%% file: hierarchical-datacubes-en-jcc.tex
\documentclass[a4 paper, 11pt,twoside]{article}
\usepackage{fullpage} 
\usepackage[lmargin=0.6in,rmargin=0.6in,tmargin=0.6in,headsep=.2in]{geometry}
\usepackage{graphicx}
\usepackage{forloop}
\usepackage{etoolbox}
\usepackage{calc}
\usepackage{wrapfig,lipsum,booktabs} 
\usepackage{xtab} 
\usepackage[dvipsnames]{xcolor}
\usepackage[normalem]{ulem}
\usepackage{sectsty} 
\allsectionsfont{\color{Brown}}
\makeatletter
\def\@seccntformat#1{\@ifundefined{#1@cntformat}%
    {\csname the#1\endcsname\;} 
    {\csname #1@cntformat\endcsname} 
}
\def\section@cntformat{\thesection.\;} 
\def\subsection@cntformat{\thesubsection.\;} 
\makeatother
\usepackage{hyperref}
\hypersetup{
    colorlinks=true,
    urlcolor=blue,
    linkcolor=black,
    citecolor = black
}

\usepackage{fancyhdr}
\fancypagestyle{first}{%
    \fancyhf{} 
    \fancyhead[L]{\includegraphics[scale=0.8]{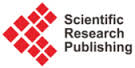}}%
    \fancyfoot[L]{{\footnotesize \textsf{DOI: \href{10.4236/jcc.2023.116004}{\color{blue}\uline{10.4236/jcc.2023.116004}} $\;$June 30, 2023}}}%
    \fancyhead[R]{{\bf\small \textsf{Journal of Computer and Communications, 2023, 11, 43-72}}\\
\href{http://www.scirp.org/journal/jcc}{\color{blue}\uline{\textsf{http://www.scirp.org/journal/jcc}}}\\
\textsf{ISSN Online:2327-5227}\\
\textsf{ISSN Print:2327-5219}}%
}
\usepackage{amsmath,amsfonts,amssymb,amsthm,epsfig,epstopdf,url,array}
 \usepackage[retainorgcmds]{IEEEtrantools}
 \usepackage{makeidx,epsfig,lscape}
 \usepackage{xcolor,pict2e}
\usepackage{upgreek}

\theoremstyle{definition}

\newtheorem{theorem}{Theorem}
\newtheorem{proposition}[theorem]{Proposition}
\newtheorem{lemma}[theorem]{Lemma}

\InputIfFileExists{tex_config/paquetages}{}{}

\InputIfFileExists{tex_config/commandes}{}{}

\InputIfFileExists{tex_config/config}{}{}

\InputIfFileExists{tex_lang/en_EN}{}{}

\begin{document}
	
\thispagestyle{first}
\vspace*{3cm}
{\noindent\huge\bf Hierarchical datacubes}\\[1cm]
{\bf\large Mickaël Martin Nevot, Sébastien Nedjar, Lotfi Lakhal}\\[0.5cm]
Laboratoire d'Informatique et Système CRNS UMR 7020, Aix-Marseille Université, France\\
E-mail: mickael.martin-nevot@lis-lab.fr, sebastien.nedjar@lis-lab.fr, lotfi.lakhal@lis-lab.fr\\
\begin{wraptable}{l}{5.1cm}
{\footnotesize
\begin{xtabular*}{0.3\textwidth}{p{5cm}}
\noindent{\bf How to cite this paper:} Martin Nevot, M. (2023) Hierarchical datacubes, Journal of Computer and Communications, {\bf 11}, 43-72.\\
\url{http://dx.doi.org/10.4236/jcc.2023.116004}\\
{\bf Received: March 31, 2023}\\
{\bf Accepted: June 27, 2023}\\
{\bf Published: June 30, 2023}\\
Copyright \copyright$\;$2023 by author(s) and Scientific Research Publishing Inc.\\
This work is licensed under the Creative Commons Attribution International License (CC BY 4.0).\\
\url{http://creativecommons.org/licenses/by/4.0/}\\
\includegraphics[width=2.5cm,height=0.72cm]{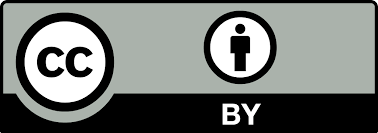}$\;$\includegraphics[width=2.5cm,height=0.75cm]{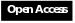}\\
{\color{white}\lipsum[1-60]}
\end{xtabular*}
}
\end{wraptable}
{\color{Brown}\rule{0.7\textwidth}{2pt}}\\[0.2cm]
{\color{Brown}\bf\large Abstract}\\
Many approaches have proposed to pre-compute data cubes in order to efficiently respond to OLAP queries in data warehouses. However, few have proposed solutions integrating all of the possible outcomes, and it is this idea that lead the integration of hierarchical dimensions into these responses. To meet this need, we propose, in this paper, a complete redefinition of the framework and the formal definition of traditional database analysis through the prism of hierarchical dimensions. After characterizing the hierarchical data cube lattice, we introduce the hierarchical data cube and its most concise reduced representation, the closed hierarchical data cube. It offers compact replication so as to optimize storage space by removing redundancies of strongly correlated data. Such data are typical of data warehouses, and in particular in video games, our field of study and experimentation, where hierarchical dimension attributes are widely represented.
\vspace{0.5cm}\\
{\color{Brown}\bf\large Keywords}\\
ROLAP cubing, Data warehouse, Datacube, Big data, Business intelligence, Hierarchical cube, Hierarchical dimensions.
\vspace{0cm}\\
{\color{Brown}\rule{0.7\textwidth}{2pt}}

\input{tex_body/article-en}

\def\cprime{$'$} \def\cprime{$'$}
\addcontentsline{toc}{section}{References}
{
	\color{black}
	
	\ifdefined\printbibliography
	    \printbibliography
	\fi
}

\end{document}

%% file: tex_config/paquetages.tex

\usepackage{lipsum}								

\usepackage{xpatch}								

\usepackage[babel]{csquotes}					

\usepackage{booktabs}							

\usepackage{amsmath}							
\usepackage{amssymb}							
\usepackage{amsfonts}							

\usepackage[sorting=none, backend=bibtex]{biblatex}			    


\usepackage{authblk}							


\usepackage[disable]{todonotes}		            

\usepackage{listings, lstautogobble}			

\usepackage[mathscr]{eucal}						

\usepackage{colortbl}							

\usepackage{amsthm}								

\usepackage[ruled]{algorithm}					
\usepackage{algorithmic}						

\usepackage{bookmark,hyperref}					

%% file: tex_config/commandes.tex

\newcommand\numberstyle[1]{%
	\footnotesize
	\color{SQLcodegray}%
	\ttfamily
	\ifnum#1<10 0\fi#1 |%
}

\makeatletter
\xpatchcmd\algorithmic
{\ALC@linenosize \arabic{ALC@line}\ALC@linenodelimiter}
{\ALC@linenosize \theALC@line\ALC@linenodelimiter}
{}{}

\newcommand\setAlgoLinenoFormat{\renewcommand*{\theALC@line}{\ifnum\value{ALC@line}<10 0\fi\arabic{ALC@line}}}
\makeatother

\setAlgoLinenoFormat


\let\oldalgorithmic\algorithmic
\renewcommand\algorithmic{\ttfamily\fontseries{l}\selectfont\oldalgorithmic}


%% file: tex_config/config.tex
\MakeAutoQuote{«}{»}

\addbibresource{biblio.bib}

\definecolor{aliceblue}{rgb}{0.94, 0.97, 1.0}
\definecolor{skyblue}{rgb}{0.53, 0.81, 0.92}
\definecolor{brightmaroon}{rgb}{0.95,0.82,0.85}

\hypersetup{%
	breaklinks=true,			
	colorlinks=true,			
	urlcolor=blue,				
	citecolor=gray,				
	linkcolor=darkgray,			
	anchorcolor=blue,			
	bookmarksopen=false,		
	linktocpage=true,			
	bookmarksnumbered,			
	pdfstartview={FitH},		
	pdftitle={Cubes de données hiérarchiques},
	pdfauthor={Mickaël Martin Nevot, Sébastien Nedjar, Lotfi Lakhal},
	pdfsubject={Article de recherche},
	pdfkeywords = {ROLAP cubing, Data warehouse, Datacube, Big data, Business intelligence, Hierarchical cube, Hierarchical dimensions},
}

\definecolor{SQLCodeGreen}{rgb}{0,0.6,0}
\definecolor{SQLcodegray}{rgb}{0.5,0.5,0.5}
\definecolor{SQLCodePurple}{HTML}{C42043}
\definecolor{SQLBackgroundcolor}{HTML}{F2F2F2}
\definecolor{SQLBookColor}{cmyk}{0,0,0,0.90}  
\color{SQLBookColor}

\lstset{upquote=true}

\lstdefinestyle{SQLStyle} {
	backgroundcolor=\color{SQLBackgroundcolor},
	commentstyle=\color{SQLCodeGreen},
	keywordstyle=\color{SQLCodePurple},
	numberstyle=\numberstyle,
	stringstyle=\color{SQLCodePurple},
	basicstyle=\footnotesize\ttfamily,
	breakatwhitespace=false,
	breaklines=true,
	captionpos=b,
	keepspaces=true,
	numbers=left,
	numbersep=10pt,
	showspaces=false,
	showstringspaces=false,
	showtabs=false,
	autogobble=true,
	literate = 	{é}{{\'e}}{1}%
				{è}{{\`e}}{1}%
				{à}{{\`a}}{1}%
				{â}{{\^a}}{1}
				{ç}{{\c{c}}}{1}%
				{œ}{{\oe}}{1}%
				{ù}{{\`u}}{1}%
				{É}{{\'E}}{1}%
				{È}{{\`E}}{1}%
				{À}{{\`A}}{1}%
				{Ç}{{\c{C}}}{1}%
				{Œ}{{\OE}}{1}%
				{Ê}{{\^E}}{1}%
				{ê}{{\^e}}{1}%
				{î}{{\^i}}{1}%
				{ï}{{\"i}}{1}
				{ô}{{\^o}}{1}%
				{û}{{\^u}}{1}%
}

\newtheoremstyle{theoremStyle}
{9pt}{9pt}				
{}						
{}						
{\bfseries}{ -}			
{ }						
{}						

\theoremstyle{theoremStyle}

\pgfdeclarelayer{background}
\pgfdeclarelayer{foreground}
\pgfsetlayers{background, main, foreground}

\usetikzlibrary{patterns}
\usetikzlibrary{automata}

%% file: tex_lang/en_EN.tex



\ifdefined\proposition\else
    \newtheorem{proposition}{Proposition}[section]
    
    \ifdefined\lemma\else
        \newtheorem{lemma}[proposition]{Lemma}
    \fi
    
    \ifdefined\corollaire\else
        \newtheorem{corollaire}[proposition]{Corollary}
    \fi
    
    \ifdefined\theorem\else
        \newtheorem{theorem}[proposition]{Theorem}
    \fi
\fi

\ifdefined\lemma\else
    \newtheorem{lemma}{Lemma}
\fi

\ifdefined\corollaire\else
    \newtheorem{corollaire}{Corollary}
\fi

\ifdefined\theorem\else
    \newtheorem{theorem}{Theorem}
\fi

\ifdefined\definition\else
    \newtheorem{definition}{Definition}[section]
\fi

\ifdefined\example\else
    \newtheorem{example}{Exemple}[section]
\fi

\ifdefined\property\else
    \newtheorem{property}{Property}[section]
\fi

\ifdefined\remarques\else
    \newtheorem{remarques}{Notes}
\fi

%% file: tex_body/article-en.tex
\tableofcontents

\section{Introduction and motivations}


Data warehouses (\cite{inmonBuildingDataWarehouse1996}), allow the storage of huge volumes of data accumulated over time in operational databases. In fact, recently, the evolution of technologies has led companies to store their data and thus preserve the "memory" of their activities. Data warehouses have been designed for this purpose. Unlike operational databases, they have some notable characteristics. First of all, they are not intended for the day-to-day management of the information system, but to provide genuine assistance in decisionmaking. Their users, the decision-makers, are therefore few in number and are interested not in the detail of the data but in general trends, according to this or that criterion, not explicitly stored. Warehouses' data are also peculiar. Often inserted in the warehouse when refreshing, these data are persistent as they will not be updated. In addition, when inserted, these data are provided with a timestamp. They are said to be historicized. From the data deposits thus constituted, it was natural to seek to make the best use of it. Here again, it is not a question of formulating classic, simple and frequent requests fetching usually some tuples \footnote{Ordered collection of the values of an indefinite number of attributes relating to the same object.}, but to carry out analyses that require aggregation of the data in order to identify major trends. Such queries are particularly expensive because they require scans of large volumes of data and they are inherently complex. However, being part of a decision support process (hence the acronym OLAP for online analytical processing\footnote{This term was defined by \cite{coddRelationalModelDatabase1990} through twelve rules to be applied to a database so that it is OLAP: \begin{enumerate} \item Multidimensional conceptual view~; \item Transparency~; \item Accessibility~; \item Constancy of response times~; \item Client/server architecture~; \item Dimension independence~; \item Management of sparse matrices~; \item Multi-user support~; \item No restrictions on inter- and intra-dimensional~ \item Intuitive data manipulation~ \item Flexible reporting~ \item Unlimited number of dimensions and unlimited number of items on dimensions. \end{enumerate}}), the formulation of these queries depends on both prior knowledge and the needs of decision-makers (\cite{chaudhuriOverviewDataWarehousing1997}, \cite{chaudhuriOverviewBusinessIntelligence2011}). OLAP is opposed to online transaction processing (OLTP). It therefore takes place on an \emph{ad hoc} basis and ideally should be interactive. However, the underlying calculation is complex and time-consuming, hence the idea of pre-calculating the results.

However, as it is not obvious how to predict the decision-makers' assumptions, the preliminary calculations must consider all possible outcomes. It is this idea that encouraged us to integrate the hierarchical dimensions into these answers (\cite{hurtadoOLAPDimensionConstraints2002}). To meet this need, we propose in this paper a complete redefinition of the framework and formal definition of traditional database analysis (\cite{kuijpersFormalAlgebraOLAP2016}), the datacube, and its closed datacube through the prism of hierarchical dimensions. Hierarchies add their own modeling complexity to those of data warehouses (\cite{malinowskiOLAPHierarchiesConceptual2004}), it is for this reason, and the consequent computation times that they induce, that few authors have been interested in this field. However, we believe that a work of framing and formal definition of hierarchical dimensions is judicious. Moreover, in general, we will consider two classes of OLAP operators. The first is dedicated to the data structure and allows its manipulation in order to retrieve remarkable information. The second is dedicated to the granularity of the data and proposes operators that aggregate and synthesise the information in one direction (roll-up), and which break them down or specify them in the other (drill down) (\cite{favreEntrepotsDonneesPour2013}). Dimensional hierarchies define the access paths in the data and allow the implementation of this second class of OLAP operators. It is also for this reason that studying the hierarchical dimensions is relevant although the volume of data generated, even greater than in the case of the classic datacube, must be understood by a particularly reduced approach to be completely usable.

Computing a hierarchical datacube, taking into account the hierarchical structure of the dimensions, can be judicious to obtain an optimized aggregation, a simplified analysis, an intuitive navigation and to help decision-making based on the hierarchies. Using the hierarchical structure of dimensions, it is possible to perform optimized aggregations by consolidating data at different levels of granularity. This allows data to be pre-aggregated at higher levels of the hierarchy, which reduces the size of the data cube and therefore improves query performance.A hierarchical datacube also helps simplify analysis by providing an aggregated view of data at different levels of granularity. The aim is to facilitate the understanding of trends and relationships between the different dimensions, avoiding the complexity of detailed data. The hierarchical structure of the dimensions in a hierarchical datacube also allows intuitive and easy navigation between the different levels of granularity.This allows users to explore data more efficiently and navigate between levels of detail and aggregations without extra effort. This can improve user experience and speed up data analysis. Finally, in many cases, the hierarchical structure of dimensions is meaningful from a decision-making perspective. By using a hierarchical datacube, it is possible to take this hierarchical structure into account to obtain more relevant decision information that is better aligned with the reality of the domain.

\section{Use cases}

\subsection{Open Match 3}

For this paper, we introduce an example applied to a video game, the one we initially studied and experimented with when researching this contribution.

\begin{figure}[htbp]
	\centering
	\includegraphics[width=240px]{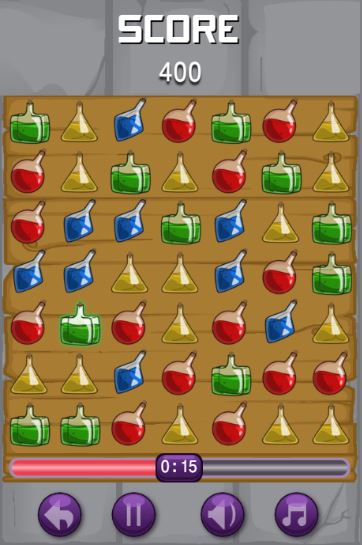}
	\caption{Video game Open Match 3}\label{fig:jeu_video_open_match_3}
\end{figure}

In this example, we will use the video puzzle-game\footnote{A video game of puzzle is a type of video game based on reflection. Its name comes from the jigsaw puzzle, a game consisting of putting pieces in order. It can be a game in which the player has to move elements in a specific way. Many of these games are called tile-matching games.} of the  Match 3\footnote{A match 3 game is a type of video game in which the player must combine three elements of the same colour or shape in order to eliminate them from the game board. These games are often considered as reflection games and are generally based on the combination of strategy and luck. Match 3 games can be played on various platforms, including computers, mobile phones, and game consoles, and they are generally very popular with different age groups. Match 3 games can be simple entertainment or they can offer increasingly complex challenges as the player progresses through the game.} type open source \footnote{The open source term designates an approach to software development (and now extends to works of the mind) whose license respects criteria precisely established by the Open Source Initiative (\url{https://opensource.org/}) and in which the source code is freely available and can be used, modified and distributed by anyone. The aim of the open source approach is to allow more people to contribute to the development and improvement of software, which can lead to increased innovation and improved quality. Open source licences prescribe the conditions under which source code can be used, modified and distributed. Some licences require that changes to the source code be published so that others can benefit from them, while other licences allow the user to keep their changes private ).} Open Match 3, a web browser-based video game \footnote{A browser-based video game is a type of online game that can be played through a web browser and does not require installation or downloading.}, developed by Giordano Ferdinandi, Stephen Surtees and Rachel Kehoe of Clockwork Chilli.

Initially, we focused on this type of video game because of its immense popularity, making the example more eloquent and its application relevant. In addition, besides its ethical side, this choice of open source allowed us to have access and to directly modify the source code of the video game in order to accommodate all the probes we wanted to listen to.

\subsection{Relation example}

\begin{figure}[htbp]
	\centering
	\begin{tikzpicture}[
		line join=bevel,
		]
		
		\node (fact) at (150pt, 200pt)
		{
			\begin{tabular}{c}
				\toprule
				$\texttt{Fact}_{\texttt{OM3}}$ \\
				\midrule
				$\texttt{RowId}$               \\
				\midrule
				$\texttt{IdPlayer}$            \\
				$\texttt{IdTurn}$              \\
				$\texttt{IdSeries}$            \\
				\hline
				$\texttt{Time}$                \\
				$\texttt{Duration}$            \\
				$\texttt{Number}$              \\
				$\texttt{Score}$               \\
				$\texttt{Shape}$               \\
				\bottomrule
			\end{tabular}
		};
		\node (dPlayer) at (150pt, 0pt)
		{
			\begin{tabular}{c}
				\toprule
				$\texttt{D}_{\texttt{Player}}$ \\ 
				\midrule
				$\texttt{IdPlayer}$            \\
				\hline
				$\texttt{Country}$             \\
				$\texttt{Region}$              \\
				$\texttt{City}$                \\	
				$\texttt{IPAddress}$           \\
				$\texttt{OS}$                  \\
				$\texttt{Browser}$             \\
				$\texttt{Lang}$                \\
				$\texttt{Player}$              \\
				\hline
				$\texttt{Location}$            \\
				\bottomrule
			\end{tabular}
		};
		\node (dTurn) at (0pt, 300pt)
		{
			\begin{tabular}{c}
				\toprule
				$\texttt{D}_{\texttt{Turn}}$ \\  
				\midrule       
				$\texttt{IdTurn}$            \\
				\hline
				$\texttt{Game}$              \\
				$\texttt{Round}$             \\
				\hline
				$\texttt{Step}$              \\
				\bottomrule
			\end{tabular}
		};
		\node (dSeries) at (300pt, 300pt)
		{
			\begin{tabular}{c}
				\toprule
				$\texttt{D}_{\texttt{Series}}$ \\ 
				\midrule
				$\texttt{IdSeries}$            \\ 
				\hline
				$\texttt{Move}$                \\
				$\texttt{Combination}$         \\
				\hline
				$\texttt{Association}$         \\
				$\texttt{Color}$               \\
				$\texttt{Size}$                \\
				\bottomrule
			\end{tabular}
		};
		
		\draw [] (fact) -- (dSeries);
		\draw [] (fact) -- (dTurn);
		\draw [] (fact) -- (dPlayer);
	\end{tikzpicture}
	\caption{Star schema of data warehouse \texttt{OM3}}\label{fig:schema_en_etoile_de_l_entrepot_om3}
\end{figure}
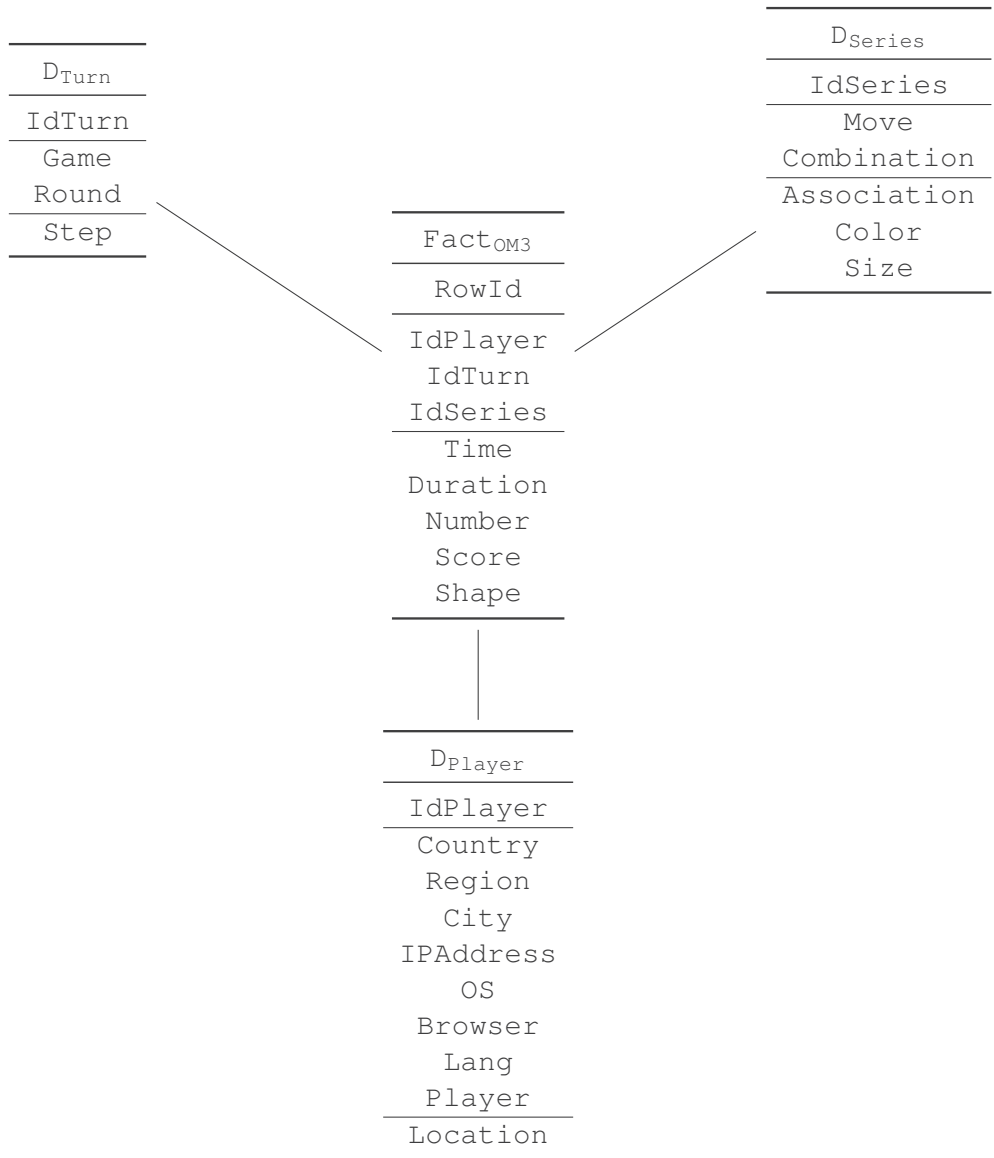

Let's consider the \texttt{OM3} data warehouse (\emph{cf.} figure \ref{fig:schema_en_etoile_de_l_entrepot_om3}) of the Open Match 3 video game (\emph{cf.} figure \ref{fig:jeu_video_open_match_3}).

A $\texttt{D}_{\texttt{Turn}}$ is identified by \texttt{IdTurn} and classified by \texttt{Game} and by \texttt{Round}. These attributes constitute a hierarchical structure, called \texttt{Step} (\emph{cf.} table \ref{tab:d_tour}).

\begin{figure}[htbp]
	\centering
	\includegraphics[width=240px]{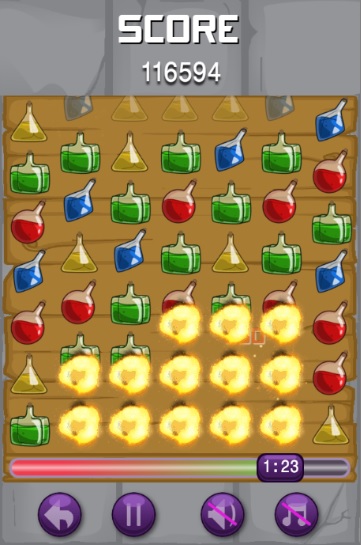}
	\caption{Video game Open Match 3, example of an explosion}\label{fig:jeu_video_open_match_3_exemple_d_explosion}
\end{figure}

A $\texttt{D}_{\texttt{Series}}$ is identified by \texttt{IdSeries} and characterized by a \texttt{Move} causing one (or more, then called a chain) \texttt{Combination} of an \texttt{Association}, characterized by a matching \texttt{Color} (among the 4) and a \texttt{Size} of at least three elements. From an \texttt{Association} of four elements or more, bonus points and special elements (explosion, rayon destructeur, etc., \emph{cf.} figure \ref{fig:jeu_video_open_match_3_exemple_d_explosion}) are granted (\emph{cf.} table \ref{tab:d_serie}).

Un $\texttt{D}_{\texttt{Player}}$ is identified by \texttt{IdPlayer} and classified by its \texttt{Location} (full). The most general location type is a \texttt{Country} (\texttt{C}), which includes a \texttt{Region} (\texttt{R}), which itself includes a \texttt{City} (\texttt{C}), specified by an \texttt{IPAddress} (\texttt{I}), itself specified by an \texttt{OS} (\texttt{O}, \emph{i.e.} an operating system), a \texttt{Browser} (\texttt{B}), then a \texttt{Lang} (\texttt{L}) thus, at last, a \texttt{Player} (\texttt{P}) name. The values of the different attributes of $\texttt{D}_{\texttt{Player}}$ are coded as follows (\emph{cf.} table \ref{tab:d_joueur}):
\begin{center}
	\begin{tabular}{ll|ll|ll|ll}
		\toprule
		\texttt{Country} & & \texttt{Region} & & \texttt{City} & & \texttt{IPAddress} & \\
		\midrule
		$France$ & $ = 1$ & $IDF$  & $ = 2$ & $Paris$     & $ = 3$  & $92.88.91.80$     & $ = 4$  \\
        &        & $PACA$ & $ = 9$ & $Marseille$ & $ = 10$ & $139.124.242.125$ & $ = 11$ \\
        &        &        &        &             &         &                   &         \\
		\bottomrule
	\end{tabular}
	\begin{tabular}{ll|ll|ll|ll}
		\toprule
		\texttt{OS} & & \texttt{Browser} & & \texttt{Lang} & & \texttt{Player} & \\
		\midrule
		$Windows$ & $ = 5$  & $Chrome$  & $ = 6$  & $fr$ & $ = 7$  & $P_1$ & $ = 8$  \\
		$Linux$   & $ = 12$ & $Opera$   & $ = 13$ & $en$ & $ = 14$ & $P_2$ & $ = 15$ \\
		$Mac\;OS$ & $ = 16$ & $Firefox$ & $ = 17$ & $es$ & $ = 18$ & $P_3$ & $ = 19$ \\
		\bottomrule
	\end{tabular}
\end{center}

The table of facts $\texttt{Fact}_{\texttt{OM3}}$ is composed of a \texttt{Time} remaining before the end of the game (naturally decreasing, but also increasing to a maximum with the \texttt{Score}, it is thus a decisional element), a \texttt{Duration} of realization, a \texttt{Number} of elements in correspondence, a \texttt{Score} (number of points won) total and a description of \texttt{Shape} ($0.5$ for horizontal, $1$ for vertical, and between these two values for mixed) for a \texttt{IdPlayer} (\texttt{IdP}), a \texttt{IdTurn} (\texttt{IdT}) and a \texttt{IdSeries} (\texttt{IdS}). The values of the different attributes $\texttt{Fact}_{\texttt{OM3}}$ are coded as follows (\emph{cf.} table \ref{tab:fait_om3}):
\begin{center}
	\begin{tabular}{ll|ll|ll}
		\toprule
		\texttt{IdP} & & \texttt{IdT} & & \texttt{IdS} & \\
		\midrule
		$P_1$ & $ = 8$  & $S_1$ & $ = 1$ & $A_1$ & $ = 1$  \\
		$P_2$ & $ = 15$ & $S_2$ & $ = 4$ & $A_2$ & $ = 4$  \\
		$P_3$ & $ = 19$ & $S_3$ & $ = 8$ & $A_3$ & $ = 11$ \\
		      &         &       &        & $A_4$ & $ = 19$ \\
	 	      &         &       &        & $A_5$ & $ = 21$ \\
		      &         &       &        & $A_6$ & $ = 23$ \\
		      &         &       &        & $A_7$ & $ = 25$ \\
		\bottomrule
	\end{tabular}
\end{center}

\begin{figure}[htbp]
	\centering
	\scriptsize
	\resizebox{1\textwidth}{!}{
		\begin{tikzpicture}[
			line join=bevel,
			]
			
			\node (nh0) at (0pt, 0pt) {$\texttt{Player}$};
			
			\node (n01) at (100pt, 0pt) {$P_1$};
			\node (n02) at (200pt, 0pt) {$P_2$};
			\node (n03) at (300pt, 0pt) {$P_3$};
			\node (nh1) at (0pt, 40pt) {$\texttt{Lang}$};
			
			\node (n11) at (100pt, 40pt) {$fr$};
			\node (n12) at (200pt, 40pt) {$en$};
			\node (n13) at (300pt, 40pt) {$es$};
			\node (nh2) at (0pt, 80pt) {$\texttt{Browser}$};
			
			\node (n21) at (100pt, 80pt) {$Chrome$};
			\node (n22) at (200pt, 80pt) {$Opera$};
			\node (n23) at (300pt, 80pt) {$Firefox$};
			\node (nh3) at (0pt, 120pt) {$\texttt{OS}$};
			
			\node (n31) at (100pt, 120pt) {$Windows$};
			\node (n32) at (200pt, 120pt) {$Linux$};
			\node (n33) at (300pt, 120pt) {$Max\;OS$};
			\node (nh4) at (0pt, 160pt) {$\texttt{IPAddress}$};
			
			\node (n41) at (100pt, 160pt) {$92.88.91.80$};
			\node (n42) at (200pt, 160pt) {$139.124.242.125$};
			\node (nh5) at (0pt, 200pt) {$\texttt{City}$};
			
			\node (n51) at (100pt, 200pt) {$Paris$};
			\node (n52) at (200pt, 200pt) {$Marseille$};
			\node (nh6) at (0pt, 240pt) {$\texttt{Region}$};
			
			\node (n61) at (100pt, 240pt) {$IDF$};
			\node (n62) at (200pt, 240pt) {$PACA$};
			\node (nh7) at (0pt, 280pt) {$\texttt{Country}$};
			
			\node (n71) at (150pt, 280pt) {$France$};
			\node (nh8) at (0pt, 320pt) {$\top_{\texttt{Player}}$};
			
			\node (n81) at (150pt, 320pt) {$ALL_{\texttt{Player}}$};
			\node (nhtop) at (0pt, 360pt) {$S_h({\texttt{D}_{\texttt{Player}}})$};
			\node (ntop) at (200pt, 360pt) {$h_\texttt{Player}$};
			
			\draw [] (n01) -- (n11);
			\draw [] (n02) -- (n12);
			\draw [] (n03) -- (n13);
			\draw [] (n11) -- (n21);
			\draw [] (n12) -- (n22);
			\draw [] (n13) -- (n23);
			\draw [] (n21) -- (n31);
			\draw [] (n22) -- (n32);
			\draw [] (n23) -- (n33);
			\draw [] (n31) -- (n41);
			\draw [] (n32) -- (n42);
			\draw [] (n33) -- (n42);
			\draw [] (n41) -- (n51);
			\draw [] (n42) -- (n52);
			\draw [] (n51) -- (n61);
			\draw [] (n52) -- (n62);
			\draw [] (n61) -- (n71);
			\draw [] (n62) -- (n71);
			\draw [] (n71) -- (n81);
			\draw [] (50pt, 0pt) -- (50pt, 360pt);
		\end{tikzpicture}
	}
	\caption{Dimension of $\texttt{D}_{\texttt{Player}}$}\label{fig:dimension_de_d_joueur}
\end{figure}
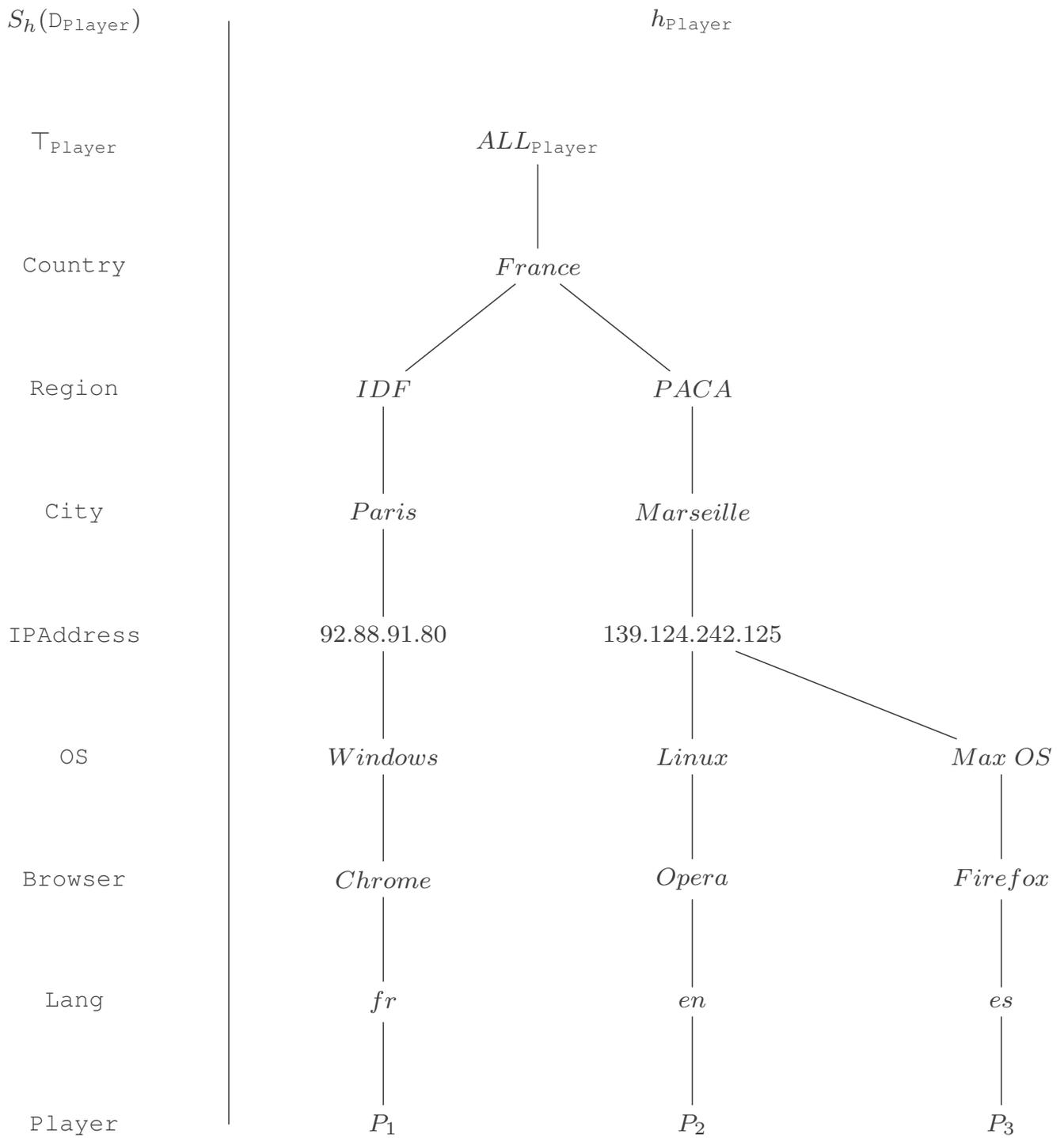

We only give a representation of a hierarchical dimension for $\texttt{D}_{\texttt{Player}}$ (\emph{cf.} figure \ref{fig:dimension_de_d_joueur}), $\texttt{D}_{\texttt{Turn}}$ and $\texttt{D}_{\texttt{Series}}$ being hierarchical dimensionswhich are much less well founded than this one. The maximum element $ALL_{\texttt{Player}}$ of this representation is defined by \ref{def:element_maximal_d_une_hierarchie} in subsection \ref{ssec:cadre_formel_d_une_dimension_hierarchique}.

\begin{table}[htbp]
	\caption{Relation of the dimension $\texttt{D}_{\texttt{Turn}}$}\label{tab:d_tour}
	\centering
	\begin{tabular}{c|cc|c}
		\toprule   
		\texttt{IdTurn} & \texttt{Game} & \texttt{Round} & \texttt{Step} \\
		\midrule
		$1$  & $1$ &      & $S_1$                    \\
		$2$  & $1$ & $2$  & $\operatorname{S_{1-1}}$ \\
		$3$  & $1$ & $3$  & $\operatorname{S_{1-2}}$ \\
		$4$  & $4$ &      & $S_2$                    \\
		$5$  & $4$ & $5$  & $\operatorname{S_{2-1}}$ \\
		$6$  & $4$ & $6$  & $\operatorname{S_{2-2}}$ \\
		$7$  & $4$ & $7$  & $\operatorname{S_{2-3}}$ \\
		$8$  & $8$ &      & $S_3$                    \\
		$9$  & $8$ & $9$  & $\operatorname{S_{3-1}}$ \\
		$10$ & $8$ & $10$ & $\operatorname{S_{3-2}}$ \\
		$11$ & $8$ & $11$ & $\operatorname{S_{3-3}}$ \\
		\bottomrule
	\end{tabular}
\end{table}

\begin{table}[htbp]
	\caption{Relation of the dimension $\texttt{D}_{\texttt{Series}}$}\label{tab:d_serie}
	\centering
	\begin{tabular}{c|cc|ccc}
		\toprule   
		\texttt{IdSeries} & \texttt{Move} & \texttt{Combination} & \texttt{Association} & \texttt{Color} & \texttt{Size} \\
		\midrule
		$1$  & $1$  &      & $A_1$                     &          &     \\
		$2$  & $1$  & $2$  & $\operatorname{A_{1-1}}$  & $red$    & $3$ \\
		$3$  & $1$  & $3$  & $\operatorname{A_{1-2}}$  & $blue$   & $3$ \\
		$4$  & $4$  &      & $A_2$                     &          &     \\
		$5$  & $4$  & $5$  & $\operatorname{A_{2-1}}$  & $green$  & $4$ \\
		$6$  & $4$  & $6$  & $\operatorname{A_{2-2}}$  & $yellow$ & $3$ \\
		$7$  & $4$  & $7$  & $\operatorname{A_{2-3}}$  & $red$    & $3$ \\
		$8$  & $4$  & $8$  & $\operatorname{A_{2-4}}$  & $blue$   & $3$ \\
		$9$  & $4$  & $9$  & $\operatorname{A_{2-5}}$  & $yellow$ & $3$ \\
		$10$ & $4$  & $10$ & $\operatorname{A_{2-6}}$  & $green$  & $4$ \\
		$11$ & $11$ &      & $A_3$                     &          &     \\
		$12$ & $11$ & $12$ & $\operatorname{A_{3-1}}$  & $green$  & $3$ \\
		$13$ & $11$ & $13$ & $\operatorname{A_{3-2}}$  & $red$    & $3$ \\
		$14$ & $11$ & $14$ & $\operatorname{A_{3-3}}$  & $red$    & $4$ \\
		$15$ & $11$ & $15$ & $\operatorname{A_{3-4}}$  & $green$  & $3$ \\
		$16$ & $11$ & $16$ & $\operatorname{A_{3-5}}$  & $yellow$ & $4$ \\
		$17$ & $11$ & $17$ & $\operatorname{A_{3-6}}$  & $yellow$ & $3$ \\
		$18$ & $11$ & $18$ & $\operatorname{A_{3-7}}$  & $yellow$ & $3$ \\
		$19$ & $19$ &      & $A_4$                     &          &     \\
		$20$ & $19$ & $20$ & $\operatorname{A_{4-1}}$  & $green$  & $3$ \\
		$21$ & $21$ &      & $A_5$                     &          &     \\
		$22$ & $21$ & $22$ & $\operatorname{A_{5-1}}$  & $red$    & $3$ \\
		$23$ & $23$ &      & $A_6$                     &          &     \\
		$24$ & $23$ & $24$ & $\operatorname{A_{6-1}}$  & $green$  & $5$ \\
		$25$ & $25$ &      & $A_7$                     &          &     \\
		$26$ & $25$ & $26$ & $\operatorname{A_{7-1}}$  & $yellow$ & $3$ \\
		$27$ & $25$ & $27$ & $\operatorname{A_{7-2}}$  & $yellow$ & $3$ \\
		$28$ & $25$ & $28$ & $\operatorname{A_{7-3}}$  & $red$    & $3$ \\
		$29$ & $25$ & $29$ & $\operatorname{A_{7-4}}$  & $red$    & $3$ \\
		$30$ & $25$ & $30$ & $\operatorname{A_{7-5}}$  & $yellow$ & $3$ \\
		$31$ & $25$ & $31$ & $\operatorname{A_{7-6}}$  & $red$    & $4$ \\
		$32$ & $25$ & $32$ & $\operatorname{A_{7-7}}$  & $green$  & $3$ \\
		$33$ & $25$ & $33$ & $\operatorname{A_{7-8}}$  & $yellow$ & $3$ \\
		$34$ & $25$ & $34$ & $\operatorname{A_{7-9}}$  & $yellow$ & $3$ \\
		$35$ & $25$ & $35$ & $\operatorname{A_{7-10}}$ & $green$  & $3$ \\
		$36$ & $25$ & $36$ & $\operatorname{A_{7-11}}$ & $green$  & $4$ \\
		$37$ & $25$ & $37$ & $\operatorname{A_{7-12}}$ & $blue$   & $3$ \\
		\bottomrule
	\end{tabular}
\end{table}

\begin{table}[htbp]
	\caption{Relation of the dimension $\texttt{D}_{\texttt{Player}}$}\label{tab:d_joueur}
	\centering
	\begin{tabular}{c|cccccccc|c}
		\toprule   
		\texttt{IdPlayer} & \texttt{C} & \texttt{R} & \texttt{C} & \texttt{I} & \texttt{O} & \texttt{B} & \texttt{L} & \texttt{P} & \texttt{Location} \\
		\midrule
		$1$  & $1$ &     &      &      &      &      &      &      & $France$          \\
		$2$  & $1$ & $2$ &      &      &      &      &      &      & $IDF$             \\
		$3$  & $1$ & $2$ & $3$  &      &      &      &      &      & $Paris$           \\
		$4$  & $1$ & $2$ & $3$  & $4$  &      &      &      &      & $92.88.91.80 $    \\
		$5$  & $1$ & $2$ & $3$  & $4$  & $5$  &      &      &      & $Windows$         \\
		$6$  & $1$ & $2$ & $3$  & $4$  & $5$  & $6$  &      &      & $Chrome$          \\
		$7$  & $1$ & $2$ & $3$  & $4$  & $5$  & $6$  & $7$  &      & $fr$              \\
		$8$  & $1$ & $2$ & $3$  & $4$  & $5$  & $6$  & $7$  & $8$  & $P_1$             \\
		$9$  & $1$ & $9$ &      &      &      &      &      &      & $PACA$            \\
		$10$ & $1$ & $9$ & $10$ &      &      &      &      &      & $Marseille$       \\
		$11$ & $1$ & $9$ & $10$ & $11$ &      &      &      &      & $139.124.242.125$ \\
		$12$ & $1$ & $9$ & $10$ & $11$ & $12$ &      &      &      & $Linux$           \\
		$13$ & $1$ & $9$ & $10$ & $11$ & $12$ & $13$ &      &      & $Opera$           \\
		$14$ & $1$ & $9$ & $10$ & $11$ & $12$ & $13$ & $14$ &      & $en$              \\
		$15$ & $1$ & $9$ & $10$ & $11$ & $12$ & $13$ & $14$ & $15$ & $P_2$             \\
		$16$ & $1$ & $9$ & $10$ & $11$ & $16$ &      &      &      & $Mac\;OS$         \\
		$17$ & $1$ & $9$ & $10$ & $11$ & $16$ & $17$ &      &      & $Firefox$         \\
		$18$ & $1$ & $9$ & $10$ & $11$ & $16$ & $17$ & $18$ &      & $es$              \\
		$19$ & $1$ & $9$ & $10$ & $11$ & $16$ & $17$ & $18$ & $19$ & $P_3$             \\
		\bottomrule
	\end{tabular}
\end{table}

\begin{table}[htbp]
	\caption{Relation of the table of facts $\texttt{Fact}_{\texttt{OM3}}$}\label{tab:fait_om3}
	\centering
	\begin{tabular}{c|ccc|ccccc}
		\toprule
		\texttt{RowId} & \texttt{IdP} & \texttt{IdT} & \texttt{IdS} & \texttt{Time} & \texttt{Duration} & \texttt{Number} & \texttt{Score} & \texttt{Shape} \\
		\midrule
		$1$ & $P_1$ & $S_1$ & $A_1$ & $6.32$  & $2.85$ & $3.5$  & $700$  & $0.5$  \\
		$2$ & $P_1$ & $S_1$ & $A_2$ & $18.9$  & $1.95$ & $3.83$ & $2300$ & $0.5$  \\
		$3$ & $P_2$ & $S_2$ & $A_3$ & $26.39$ & $1.7$  & $3.43$ & $2400$ & $0.71$ \\
		$4$ & $P_2$ & $S_2$ & $A_4$ & $4.1$   & $2.07$ & $3$    & $300$  & $0.5$  \\
		$5$ & $P_2$ & $S_2$ & $A_5$ & $7.38$  & $3.68$ & $3$    & $600$  & $0.5$  \\
		$6$ & $P_3$ & $S_3$ & $A_6$ & $2.14$  & $2.15$ & $3$    & $300$  & $1$    \\
		$7$ & $P_3$ & $S_3$ & $A_7$ & $56.04$ & $2.25$ & $3.17$ & $3800$ & $0.5$  \\
		\bottomrule 
	\end{tabular}
\end{table}

\section{Hierarchical dimension}

\subsection{Types of hierarchies}

Dimensional hierarchies are formed by the set of attributes of a dimensional relation having a hierarchical relationship ($a \in a'$). They correspond to relations of type $1$ to several and offer the possibility of using the class of roll up and drill down (\cite{favreEntrepotsDonneesPour2013}).

The literature provides an overview of the different types of hierarchies (\cite{inmonBuildingDataWarehouse1996}) that we propose to present.

\subsubsection{Simple Hierarchy}\index{Simple Hierarchy}

Simple hierarchies have levels that can be considered as lists and their instances form a tree. These hierarchies can be symmetrical if the tree of instances is balanced or asymmetrical if it is not.

\subsubsection{Strict Hierarchy and Non-strict Hierarchy}

\index{Hiérarchie stricte} In the framework of a strict hierarchy, the generalisation of a value at a given level only leads to, at most, one value.

\begin{example}
	The month-level generalisation of the date July, 14th 1789 is July.
\end{example}

\index{Hiérarchie non-stricte} In practice, there are many situations where this type of hierarchy is too restrictive. Typically, a video game can be of several game types. Such hierarchies are called non-strict.

\subsubsection{Multiple Hierarchy}\index{Multiple Hierarchy}

Multiple hierarchies make it possible to model situations where the different levels are no longer lists but directed graphs without cycles.

\begin{example}
	Let us consider a time dimension. There are two ways of generalising the day level: on the week level or the month level, although these two levels are not comparable (\emph{i.e.} one is not a generalisation of the other).
\end{example}

\subsubsection{Parallel Hierarchies}

For dimensions with only one attribute, there can only be one hierarchy per dimension. In practice, it is common for a single dimension to be composed of several attributes useful for the analysis.

If these attributes have an associated hierarchy, the dimensional hierarchy is said to be parallel and can be considered as a special case of multiple hierarchies. A parallel hierarchy can be composed of other types of hierarchies. Two categories of parallel hierarchies are generally distinguished: 
\begin{itemize}
	\item independent parallel hierarchy: there is no common level between the different hierarchies composing it;
	\item parallel hierarchy: there are common levels between the different hierarchies composing it; the hierarchy is said to be independent.
\end{itemize}

\subsection{Formal framework of a hierarchical dimension}\label{ssec:cadre_formel_d_une_dimension_hierarchique}

Let $r$ be a relation of schema $\mathcal{R}$. In addition to the attributes of $\mathcal{R}$, we will consider an additional attribute $RowId$ qui which is implicit. The values of this attribute serve as a unique identifier for each tuple and are assigned at the time of their insertion. We note $t_x$ the tuple such that $t[RowId] = x$.

The attributes of $\mathcal{R}$ are divided into two sets:
\begin{enumerate}
	\item $ \mathcal{D} $ is the set of attributes of hierarchical dimensions, also called hierarchical categories. Moreover, the attributes of $\mathcal{D}$ are ordered.
	\item  $\mathcal{M}$ is the set of measured attributes, on which the aggregative or statistical functions are applied.
\end{enumerate}

We simply call dimension a hierarchical dimension attribute. Likewise, we call hierarchy a hierarchy of values.

\begin{definition}[Dimension]\index{Dimension}\index{Attribut dimension}
	In $r$, we call a dimension, a dimensional attribute, or a category, $d_i$. $\forall d_i \in \mathcal{D}$, $r(d_i)$ is the projection of $r$ onto $d_i$.
	
	A dimension is characterised by a structure (\emph{cf.} definition \ref{def:structure_d_une_dimension}) and a hierarchy (\emph{cf.} definition \ref{def:hierarchie}).
	
	We call $\mathcal{D}$ the set of dimensions $d_i$ of $r$.
\end{definition}

\begin{example}
	Let's consider the data warehouse \texttt{OM3}. Its set of dimensions $\mathcal{D}$ is:
	\[\mathcal{D} = \{\texttt{D}_{\texttt{Player}}, \texttt{D}_{\texttt{Turn}}, \texttt{D}_{\texttt{Series}}\}\]
\end{example}

\begin{definition}[Structure of a dimension]\index{Structure of a dimension}\label{def:structure_d_une_dimension}
	For each dimension $d_i \in \mathcal{D}$, we define the structure of a dimension, or schema, $S_h$ as follows: $\forall d_i \in \mathcal{D}, S_h(d_i)$ is the structure of the dimension $d_i$.
\end{definition}

\begin{example}
	Let's consider the set of hierarchical dimension structures:
	\begin{align*}
		S_h({\texttt{D}_{\texttt{Player}}}) &= (\top_{\texttt{Player}}, \texttt{Country},  \texttt{Region}, \texttt{City}, \texttt{IPAddress}, \\
		&\phantom{{}=1} \texttt{OS}, \texttt{Browser}, \texttt{Lang}, \texttt{Player}) \\
		S_h({\texttt{D}_{\texttt{Turn}}}) &= (\top_{\texttt{Turn}}, \texttt{Game},  \texttt{Round}) \\
		S_h({\texttt{D}_{\texttt{Series}}}) &= (\top_{\texttt{Series}}, \texttt{Move}, \texttt{Combination}) \\
	\end{align*}
\end{example}

\begin{definition}[Hierarchy]\index{Hierarchy}\label{def:hierarchie}
	For each dimension $d_i \in \mathcal{D}$, we define $h_i$ the hierarchy associated with that dimension, which we will simply call hierarchy.
\end{definition}

\begin{definition}[Level of a hierarchy]\index{Level of a hierarchy}
	For each hierarchy $h_i$, we define $h_i(e)$, the level, or tier, or echelon, of this hierarchy.
\end{definition}

\begin{definition}[Maximum element of a hierarchy]\index{Maximum element of a hierarchy}\label{def:element_maximal_d_une_hierarchie}
	A hierarchy $h_i$ associated with the dimension $d_i \in \mathcal{D}$ contains a maximum element denoted $\top_i$ defined by the value $ALL_i$. This value takes the idea of the $ALL$ from \cite{grayDataCubeRelational1997}, which represents the generalization of all the values of the domain of a dimension.
\end{definition}

\begin{example}
	Considering the hierarchy $h_\texttt{Player}$ of the dimension $\texttt{D}_{\texttt{Player}}$ (\emph{cf.} figure \ref{fig:dimension_de_d_joueur}), the value of its maximum element is $ALL_\texttt{Player}$.
\end{example}

\begin{definition}[Depth of a hierarchy]\index{Depth of a hierarchy}
	For each hierarchy $h_i$, we define its $Depth$ as follows: $\forall h_i \in d_i, d_i \in \mathcal{D}, Prof(h_i)$ is the depth of the hierarchy $h_i$, \emph{i.e.} its number of levels.
\end{definition}

\begin{example}
	Considering the hierarchy $h_\texttt{Player}$ of the dimension $\texttt{D}_{\texttt{Player}}$ (\emph{cf.} figure \ref{fig:dimension_de_d_joueur}), $Depth(h_\texttt{Player}) = 9$.
\end{example}

\begin{definition}[Domain of a dimension]\index{Domain of a dimension}
	We define the domain of a dimension $Dom$ as follows: $\forall d_i \in \mathcal{D}, Dom(d_i)$ is the domain of dimension $d_i$.
\end{definition}

\begin{definition}[Domain of a level of a dimension]\index{Domaine d'un niveau d'une dimension}\index{Empreinte dimensionnelle}
	We define the domain of a level of a dimension, or dimensional footprint, $Dom$ as follows:
	\begin{itemize}
		\item $\forall d_i \in \mathcal{D}, \forall e \in S_h(d_i), Dom(d_i, e)$ is the domain of the dimension $d_i$ for the level $e$, \emph{i.e.} $Dom(d_i, e)$ corresponds to the set of possible values for the level $e$ of the hierarchy $h_i$~;
		\item $\forall d_i \in \mathcal{D}, Dom(h_i) = {\bigcup\limits_{e \in S_h(d_i)} Dom(h_i, e)}$.
	\end{itemize}
\end{definition}

\begin{definition}[Dimensional table]\index{Dimensional table}\
	In the "star" data model used in data warehouses (or datamart), each hierarchy can be represented by a table, of dimension $\texttt{D}_{{d_i}}$ corresponding to the dimension eponymous (the two concepts are mingled), relational schema $S_h(d_i)$.
\end{definition}

\begin{definition}[Dimensional tuple]\index{Dimensional tuple}	
	For each $x \in Dom(d_i)$, we define the dimensional tuple $t_x$ corresponding to the value $x$ in the table $\texttt{D}_{d_i}$ is made up of the list of ancestors of $x$ in $h_i$.
	
	Likewise, for each $x \in Dom(d_i, e)$, we define the dimensional tuple $t_x$ corresponding to the value $x$ for the level $e$ of the hierarchy $h_i$ in the table $\texttt{D}_{d_i}$ is also made up of the list of ancestors of $x$ in $h_i$.
	
	We simply call tuple a dimensional tuple.
\end{definition}

\begin{example}
	For $x = P_2$, the tuple $t_x$ corresponding in the relation $\texttt{D}_{\texttt{Player}}$ is:
	\begin{align*}
		t_x &= (France, PACA, Marseille, 139.124.242.125, Linux, Opera, en, P_2)
	\end{align*}
	
	For $x = 139.124.242.125$, the tuple $t_x$ corresponding in the relation $\texttt{D}_{\texttt{Player}}$ is:
	\begin{align*}
		t_x &= (France, PACA, Marseille, 139.124.242.125, \\
		&\phantom{{}= ( } NULL, NULL, NULL, NULL) \\
	\end{align*}
\end{example}

\subsection{Formal definition of a hierarchical dimension}

\subsubsection{Multidimensional space}

\begin{definition}[Multidimensional space]\index{Multidimensional space}
	The multidimensional space of $r$ is noted and redefined with hierarchical dimensions as follows:
	\[Space(r) = \{\times_{d_i \in \mathcal{D}} Dom(h_i) \cup \{ALL_i, \dotsc, ALL_j\})\} \cup \{(\emptyset, \dotsc, \emptyset)\}\] 
	Where $\times$ symbolizes the Cartesian product, and $(\emptyset, \dotsc, \emptyset)$ the combination of empty values. is a tuple and represents a multidimensional pattern.
	
	The value of each maximal element $ALL_i$ of a hierarchy $h_i$ is naturally contained in the hierarchy, in the associated domain $Dom(h_i)$, and thus also in $Space(r)$.
\end{definition}

\begin{example}
	The multidimensional space of the \texttt{OM3}, data warehouse, and its table of facts (\emph{cf.} tableau \ref{tab:fait_om3}), is illustrated in  table \ref{tab:espace_multidimensionnel_hierarchique}. For the sake of clarity and conciseness, not all tuples in the multidimensional space are represented (there would be several thousand), and special values are abbreviated: $ALL_{\texttt{Player}}$ to $ALL_{\texttt{P}}$, $ALL_{\texttt{Turn}}$ to $ALL_{\texttt{T}}$ and $ALL_{\texttt{Series}}$ to $ALL_{\texttt{S}}$.
	
	Tuples identified by values $1$, \ldots, $13$ et $31$ are possible tuples for $r$ (because all their values are real), even if the tuple identified by value $31$ is not explicitly in the relation $r$. In this tuple, the symbol $\emptyset$ means "empty value". The other tuples (identifiers $14$, \ldots, $30$) cannot be tuples of $r$ because they contain at least one $ALL_i$ value. They therefore convey information at a more aggregated level of detail than the previous ones.
\end{example}

\begin{table}[htbp]
	\caption{Multidimensional space of the data warehouse \texttt{OM3}}\label{tab:espace_multidimensionnel_hierarchique}
	\centering
	\begin{tabular}{c|ccc}
		\toprule
		\texttt{RowId} & \texttt{IdP} & \texttt{IdT} & \texttt{IdS} \\
		\midrule
		$1$    & $France$           & $S_1$                    & $A_1$                     \\
		$2$    & $France$           & $S_1$                    & $\operatorname{A_{1-1}}$  \\
		$3$    & $France$           & $S_1$                    & $\operatorname{A_{1-2}}$  \\
		$4$    & $France$           & $S_1$                    & $A_2$                     \\
		\ldots & \ldots             & \ldots                   & \ldots                    \\
		$5$    & $France$           & $\operatorname{S_{1-1}}$ & $A_1$                     \\
		\ldots & \ldots             & \ldots                   & \ldots                    \\
		$6$    & $IDF$              & $S_1$                    & $A_1$                     \\
		\ldots & \ldots             & \ldots                   & \ldots                    \\
		$7$    & $es$               & $\operatorname{S_{3-3}}$ & $\operatorname{A_{7-12}}$ \\
		$8$    & $P_1$              & $S_1$                    & $A_1$                     \\
		\ldots & \ldots             & \ldots                   & \ldots                    \\
		$9$    & $P_1$              & $S_1$                    & $A_2$                     \\
		\ldots & \ldots             & \ldots                   & \ldots                    \\
		$10$   & $P_3$              & $S_3$                    & $A_7$                     \\
		\ldots & \ldots             & \ldots                   & \ldots                    \\
		$11$   & $France$           & $S_1$                    & $ALL_{\texttt{S}}$        \\
		\ldots & \ldots             & \ldots                   & \ldots                    \\
		$12$   & $PACA$             & $\operatorname{S_{3-3}}$ & $ALL_{\texttt{S}}$        \\
		$13$   & $Marseille$        & $S_1$                    & $ALL_{\texttt{S}}$        \\
		\ldots & \ldots             & \ldots                   & \ldots                    \\
		$14$   & $P_1$              & $S_1$                    & $ALL_{\texttt{S}}$        \\
		\ldots & \ldots             & \ldots                   & \ldots                    \\
		$15$   & $France$           & $ALL_{\texttt{T}}$       & $A_1$                     \\
		$16$   & $France$           & $ALL_{\texttt{T}}$       & $\operatorname{A_{1-1}}$  \\
		\ldots & \ldots             & \ldots                   & \ldots                    \\
		$17$   & $fr$               & $ALL_{\texttt{T}}$       & $\operatorname{A_{7-12}}$ \\
		$18$   & $P_1$              & $ALL_{\texttt{T}}$       & $A_1$                     \\
		\ldots & \ldots             & \ldots                   & \ldots                    \\
		$19$   & $ALL_{\texttt{P}}$ & $S_1$                    & $A_1$                     \\
		$20$   & $ALL_{\texttt{P}}$ & $S_1$                    & $\operatorname{A_{1-1}}$  \\
		\ldots & \ldots             & \ldots                   & \ldots                    \\
		$21$   & $France$           & $ALL_{\texttt{T}}$       & $ALL_{\texttt{S}}$        \\
		$22$   & $IDF$              & $ALL_{\texttt{T}}$       & $ALL_{\texttt{S}}$        \\
		$23$   & $Paris$            & $ALL_{\texttt{T}}$       & $ALL_{\texttt{S}}$        \\
		\ldots & \ldots             & \ldots                   & \ldots                    \\
		$24$   & $ALL_{\texttt{P}}$ & $S_1$                    & $ALL_{\texttt{S}}$        \\
		\ldots & \ldots             & \ldots                   & \ldots                    \\
		$25$   & $ALL_{\texttt{P}}$ & $ALL_{\texttt{T}}$       & $A_7$                     \\
		\ldots & \ldots             & \ldots                   & \ldots                    \\
		$26$   & $ALL_{\texttt{P}}$ & $ALL_{\texttt{T}}$       & $ALL_{\texttt{S}}$        \\
		$27$   & $\emptyset$        & $\emptyset$              & $\emptyset$               \\
		\bottomrule
	\end{tabular}
\end{table}

\subsubsection{Specialisation orders}

The multidimensional space of $r$ is structured by the specialisation relation between tiles. This order was originally introduced by \cite{mitchellGeneralizationSearch1982}, \cite{mitchellMachineLearning1997} in the context of concept learning.

The ordered set $CL(r) = \langle Space (r), \preceq_s \rangle$ is a complete lattice called cube lattice.

\begin{definition}[Intradimensional specialization order]\index{Intradimensional specialization order}
	Let's consider $x, y \in Dom(h_i)$:
	\[x \preceq_{d_i} y \Leftrightarrow x \text{ is an ancestor of } y \text{ on the hierarchy } h_i \text{.}\]
\end{definition}

\begin{definition}[Multidimensional specialization order]\index{Multidimensional specialization order}
	Let's consider $t, t' \subseteq$ $Space(r)$, we define the order relation $\preceq_s$ as follows:
	\[t \preceq_s t' \Leftrightarrow \forall d_i \in \mathcal{D}, t[d_i] \preceq_{d_i} t'[d_i]\]
\end{definition}

\subsubsection{Functions}

\begin{definition}[Dimensional $Attribute$ function]\index{Dimensional Attribute function}
	For a tuple $t_x$ de $d_i$, the function $Attribute_{d_i}$ returns the set of attributes whose value is different from $NULL$.
	
	Let's consider $t_x$ a tuple of $d_i$:
	\[Attribute_{d_i}(t_x) = \{A \in t_x \mid t_x[A] \neq NULL\}\]
\end{definition}

\begin{example}
	For the dimension $\texttt{D}_{\texttt{Player}}$:
	\begin{align*}
		\text{If } t_x &= (France, PACA, Marseille, 139.124.242.125, \\
		&\phantom{{}= ( } Linux, Opera, en, P_2) \\
		\text{Then } Attribute_{\texttt{D}_{\texttt{Player}}}(t_x) &= \{France, PACA, Marseille, 139.124.242.125, \\
		&\phantom{{}= \{} Linux, Opera, en, P_2\} \\
		\text{If } t_y &= (France, PACA, Marseille, 139.124.242.125, \\
		&\phantom{{}= ( } NULL, NULL, NULL, NULL) \\
		\text{Then } Attribute_{\texttt{D}_{\texttt{Player}}}(t_y) &= \{France, PACA, Marseille, 139.124.242.125\} \\
	\end{align*}
\end{example}

\subsubsection{Operators}

\begin{definition}[Dimensional $min$/$max$ operators]\index{Dimensional min operator}\index{Dimensional max operator}
	We define the operators $min$ and $max$ as follows:
	\begin{align*}
		min_{\preceq_{d_i}}(d_i) &= \{x \in Dom(h_i, e) \mid \nexists e' \in S_h(d_i) : e' < e\} \\
		max_{\preceq_{d_i}}(d_i) &= \{x \in Dom(h_i, e) \mid \nexists e' \in S_h(d_i) : e' > e\}
	\end{align*}
	
	Thus, $min_{\preceq_{d_i}}(d_i)$ represents the set of possible values of the highest (general) level of the hierarchical dimension structure. In a similar way, $max_{\preceq_{d_i}}(d_i)$ represents the set of possible values of the lowest (specific) level of the hierarchical dimension structure.
\end{definition}

\begin{example}
	For the dimension $\texttt{D}_{\texttt{Player}}$:
	\begin{align*}
		min_{\preceq_{\texttt{D}_{\texttt{Player}}}}(\texttt{D}_{\texttt{Player}}) = \{France\} \\
		max_{\preceq_{\texttt{D}_{\texttt{Player}}}}(\texttt{D}_{\texttt{Player}}) = \{P_1, P_2, P_3\}
	\end{align*}
\end{example}

The $min$/$max$ operators can be overloaded in order to be applied to a set of tuples $T$ of dimension $d_i$, noted respectively $min_{\preceq_s}(T)$ and $max_{\preceq_s}(T)$, by being implemented as follows:
\begin{align*}
	min_{\preceq_{d_i}}(T) &= \forall x \in Dom(h_i, e), t_x \mid \nexists e' \in S_h(d_i) : e' < e \\
	max_{\preceq_{d_i}}(T) &= \forall x \in Dom(h_i, e), \{t_x \mid \nexists e' \in S_h(d_i) : e' > e\}
\end{align*}

\begin{example}
	For the dimension $\texttt{D}_{\texttt{Player}}$, if $T = \{t_x, t_y\}$ with:
	\begin{align*}
		t_x &= (France, PACA, Marseille, 139.124.242.125, \\
		&\phantom{{}= ( } Linux, Opera, en, P_2) \\
		t_y &= (France, PACA, Marseille, 139.124.242.125, \\
		&\phantom{{}= ( } Mac\;OS, Firefox, es, P_3)
	\end{align*}
	So, we have:
	\begin{align*}
		min_{\preceq_{\texttt{D}_{\texttt{Player}}}}(T) &= (France, NULL, NULL, NULL, \\
		&\phantom{{}= ( } NULL, NULL, NULL, NULL) \\
		max_{\preceq_{\texttt{D}_{\texttt{Player}}}}(T) &= \{(NULL, NULL, NULL, NULL, \\
		&\phantom{{}= \{( } NULL, NULL, NULL, P_2), \\
		&\phantom{{}= \{} (NULL, NULL, NULL, NULL, \\
		&\phantom{{}= \{( } NULL, NULL, NULL, P_3)\}
	\end{align*}
\end{example}

\begin{definition}[Generalized $min$/$max$ operators on the cube lattice]\index{Generalized min operators on the cube lattice}\index{Generalized max operators on the cube lattice}	
	We generalize the $min$ operator on the cube lattice as follows:
	\[
	\left\lbrace
	\begin{array}{l}
		\forall T \subseteq CL(r), min_{\preceq_s}(T) = \{t \in T \mid \nexists t' \in T : t' \preceq_s t\} \\
		\forall d_i \in \mathcal{D}, min_{\preceq_s}(T[d_i]) = t[d_i] \in T[d_i] \mid \nexists t'[d_i] \in T[d_i] : t'[d_i] \preceq_s t[d_i] \\
	\end{array}
	\right.
	\]
	Likewise, we generalize the $max$ operator on the cube lattice as follows:
	\[
	\left\lbrace
	\begin{array}{l}
		\forall T \subseteq CL(r), max_{\preceq_s}(T) = \{t \in T \mid \nexists t' \in T: t \preceq_s t' \} \\
		\forall d_i \in \mathcal{D}, max_{\preceq_s}(T[d_i]) = \{t[d_i] \in T[d_i] \mid \nexists t'[d_i] \in T[d_i] : t[d_i] \preceq_s t'[d_i] \} \\
	\end{array}
	\right.
	\]
\end{definition}

\begin{definition}[Dimensional Sum operator]\index{Dimensional Sum operator}
	Let's consider $d_i \in \mathcal{D}$, we define the Dimensional Sum $+_{d_i}$ of $d_i$ as follows: $\forall x, y \in Dom(h_i), x +_{d_i} y$ is the nearest (small) common ancestor to $x$ and $y$ in $h_i$. In other words:
	\[
	\left.
	\begin{array}{l}
		\forall x \in Dom(h_i, f), \forall y \in Dom(h_i, g) \\
		x +_{d_i} y = z \in Dom(h_i, e) \mid z \preceq_{d_i} x \text{ and } z \preceq_{d_i} y \mid \nexists e' \in S_h(d_i) : e' > e
	\end{array}
	\right.
	\]
\end{definition}

\begin{example}
	For the dimension $\texttt{D}_{\texttt{Player}}$, if $x = Paris$ and $y = Marseille$ then $x +_{\texttt{D}_{\texttt{Player}}} y = France$.
\end{example}

The Dimensional Sum operator can be overloaded to be applied to all tuples $t_x$ and $t_y$ of a dimension $d_i$, noted $t_x +_{d_i} t_y$, by being implemented as follows:
\[
\begin{array}{l}
	\forall x \in Dom(h_i, f), \forall y \in Dom(h_i, g), \\
	\exists z \in Dom(h_i, e), e < f \text{ and } e < g, t_z = t_x +_{d_i} t_y = t_{x +_{d_i} y}
\end{array}
\]

\begin{example}
	For the dimension $\texttt{D}_{\texttt{Player}}$:
	\begin{align*}
		\text{If } x &= Paris, \\
		t_x &= (France, IDF, Paris, NULL, NULL, NULL, NULL, NULL) \\
		\text{And } y &= Marseille, \\
		t_y &= (France, PACA, Marseille, NULL, \\
		&\phantom{{}= ( } NULL, NULL, NULL, NULL) \\
		\text{Then } z &= France, \\
		t_z &= t_x +_{d_i} t_y \\
		&= (France, NULL, NULL, NULL, NULL, NULL, NULL, NULL) \\
	\end{align*}
\end{example}

\begin{definition}[Generalized Sum operator on the cube lattice]\index{Generalized Sum operator on the cube lattice}
	We generalize the Sum operator on the cube lattice as follows:
	\[
	\left\lbrace
	\begin{array}{l}
		\forall u, v \in CL(r), \{z = u + v\} \\
		\forall d_i \in \mathcal{D}, z[d_i] = u[d_i] +_{di} v[d_i]
	\end{array}
	\right.
	\]
	$z$ is the sum of tuples $u$ and $v$.
\end{definition}

\begin{definition}[Dimensional Product operator]\index{Dimensional Product operator}
	Let's consider $d_i \in \mathcal{D}$, we define the Dimensional Product $\bullet_{d_i}$ of $d_i$ as follows: $\forall x, y \in Dom(h_i), x \bullet_{d_i} y$ is the set of nearest common descendants of $x$ and $y$ in $h_i$. In other words:
	\[
	\left.
	\begin{array}{l}
		\forall f, g < Prof(h_i), \forall x \in Dom(h_i, f), \forall y \in Dom(h_i, g), \\
		x \bullet_{d_i} y = \{z \in Dom(h_i, e) \mid x \preceq_{d_i} z \text{ et } y \preceq_{d_i} z \mid \nexists e' \in S_h(d_i) : e' < e\}
	\end{array}
	\right.
	\]
	
	If $x$ and $y$ have no common descendants in $h_i$, then:
	\[x \bullet_{d_i} y = \{\emptyset\}\]
\end{definition}

\begin{example}
	For the dimension $\texttt{D}_{\texttt{Player}}$:
	\begin{align*}
		\text{If } x = Paris \text{ and } y = 92.88.91.80, x \bullet_{\texttt{D}_{\texttt{Player}}} y &= \{Windows\} \\
		\text{If } x = Marseille \text{ and } y = 139.124.242.125, x \bullet_{\texttt{D}_{\texttt{Player}}} y &= \{Linux, Mac\;OS\} \\
		\text{If } x = Paris \text{ and } y = Marseille, x \bullet_{\texttt{D}_{\texttt{Player}}} y &= \{\emptyset\}
	\end{align*}
\end{example}

The Dimensional Product operator can be overloaded to be applied to all tuples $t_x$ and $t_y$ of a dimension $d_i$, noted $t_x \bullet_{d_i} t_y$, by being implemented as follows:
\[
\left.
\begin{array}{l}
	\forall f, g < Prof(h_i), \forall x \in Dom(h_i, f), \forall y \in Dom(h_i, g), \\
	t_z = t_x \bullet_{d_i} t_y = \left\lbrace
	\begin{array}{l}
		\{t_{x \bullet_{d_i} y}\} \text{ if } \exists z \in Dom(h_i, e), f < e \text{ and } g < e \\
		\{(\emptyset, \dotsc, \emptyset)\} \text{ otherwise.} \\
	\end{array} \right. 
\end{array}
\right.
\]

\begin{example}
	For the dimension $\texttt{D}_{\texttt{Player}}$:
	\begin{align*}
		\text{If } x &= Marseille, \\
		t_x &= (France, PACA, Marseille, NULL, \\
		&\phantom{{}= ( } NULL, NULL, NULL, NULL) \\
		\text{And } y &= 139.124.242.125, \\
		t_y &= (France, PACA, Marseille, 139.124.242.125, \\
		&\phantom{{}= ( } NULL, NULL, NULL, NULL) \\
		\text{Then } z &= \{Linux, Mac\;OS\}, \\
		t_z &= t_x \bullet_{d_i} t_y \\
		&= \{(France, PACA, Marseille, 139.124.242.125, \\
		&\phantom{{}= \{( } Linux, NULL, NULL, NULL) \\
		&\phantom{{}= \{} (France, PACA, Marseille, 139.124.242.125, \\
		&\phantom{{}= \{( }Mac\;OS, NULL, NULL, NULL)\}
	\end{align*}
\end{example}

\begin{definition}[Generalized Product operator on the cube lattice]\index{Generalized Product operator on the cube lattice}
	We generalize the Product operator on the cube lattice as follows:
	\[
	\left\lbrace
	\begin{array}{l}
		\forall u, v \in CL(r), \{z = u \bullet v\} \\
		\forall d_i \in \mathcal{D}, \{z[d_i] = u[d_i] \bullet_{d_i} v[d_i]\}
	\end{array}
	\right.
	\]
	$z$ is the product of tuples $u$ and $v$.
\end{definition}

\begin{definition}[Dimensional Semi-product operator]\index{Dimensional Semi-Product operator}
	Let's consider $d_i \in \mathcal{D}$, we define the dimensional Semi-product $\odot_{d_i}$ of $d_i$ as follows: $\forall x, y \in Dom(h_i), x \odot_{d_i} y$ is the set of nearest descendants of $x$ and $y$ in $h_i$. In other words:
	\[
	\left.
	\begin{array}{l}
		\forall f < Prof(h_i), \forall x, y \in Dom(h_i, f), \\
		x \odot_{d_i} y = \{z \in Dom(h_i, e) \mid x \preceq_{d_i} z \text{ et } y \preceq_{d_i} z \mid \nexists e' \in S_h(d_i) : e' < e\} \\
	\end{array}
	\right.
	\]
	
	If $x$ and $y$ do not have the same level, or if they do not have descendants, in $h_i$, then:
	\[x \odot_{d_i} y = \{\emptyset\}\]
\end{definition}

\begin{example}
	For the dimension $\texttt{D}_{\texttt{Player}}$:
	\begin{align*}
		\text{If } x = Linux \text{ and } y = Max\;OS, x \odot_{\texttt{D}_{\texttt{Player}}} y &= \{Opera, Firefox\} \\
		\text{If } x = Marseille \text{ and } y = Max\;OS, x \odot_{\texttt{D}_{\texttt{Player}}} y &= \{\emptyset\}
	\end{align*}
\end{example}

The Dimensional Semi-product operator can be overloaded to be applied to all tuples $t_x$ and $t_y$ of a dimension $d_i$, noted $t_x \odot_{d_i} t_y$, by being implemented as follows:
\[
\left.
\begin{array}{l}
	\forall f < Prof(h_i), \forall x, y \in Dom(h_i, f), \forall z \in Dom(h_i, e), \\
	t_z = t_x \odot_{d_i} t_y = \{t_{x \odot_{d_i} y} \mid e > f \mid \nexists e' \in S_h(d_i) : e' < e\}
\end{array}
\right.
\]

\[
\left.
\begin{array}{l}
	\forall x, y \in Dom(h_i, f), \\
	t_z = t_x \odot_{d_i} t_y = \left\lbrace
	\begin{array}{l}
		\{t_{x \odot_{d_i} y}\} \text{ if } \exists z \in Dom(h_i, e), f < e \\
		\{(\emptyset, \dotsc, \emptyset)\} \text{ otherwise.} \\
	\end{array} \right. 
\end{array}
\right.
\]

\begin{example}
	For the dimension $\texttt{D}_{\texttt{Player}}$:
	\begin{align*}
		\text{If } x &= Linux, \\
		t_x &= (France, PACA, Marseille, 139.124.242.125, \\
		&\phantom{{}= ( }Linux, NULL, NULL, NULL) \\
		\text{And } y &= Mac\;OS, \\
		t_y &= (France, PACA, Marseille, 139.124.242.125, \\
		&\phantom{{}= ( }Mac\;OS, NULL, NULL, NULL) \\
		\text{Then } z &= \{Opera, Firefox\}, \\
		t_z &= t_x \odot_{d_i} t_y \\
		&= \{(France, PACA, Marseille, 139.124.242.125, \\
		&\phantom{{}= \{( }Linux, Opera, NULL, NULL) \\
		&\phantom{{}= \{} (France, PACA, Marseille, 139.124.242.125, \\ 
		&\phantom{{}= \{( }Mac\;OS, Firefox, NULL, NULL)\}
	\end{align*}
\end{example}

\begin{definition}[Generalized Semi-product operator on the cube lattice]\index{Generalized Semi-product operator on the cube lattice}
	We generalize the Semi-Product operator on the cube lattice as follows:
	\[
	\left\lbrace
	\begin{array}{l}
		\forall u, v \in CL(r), \{z = u \odot v\} \\
		\forall d_i \in \mathcal{D}, \{z[d_i] = u[d_i] \odot_{d_i} v[d_i]\}
	\end{array}
	\right.
	\]
	$z$ is the product of tuples $u$ and $v$.
\end{definition}

\subsubsection{Characterization of the hierarchical cube lattice}

By endowing the multidimensional space $Space(r)$ of $r$ with orders of specialization and using the operators, notably Sum and Product, we propose an algebraic structure called Hierarchical Datacube Lattice, or Hierarchical Cube Lattice, or simply Cube Lattice, which sets a theoretical and general framework for multidimensional database mining. It is easily transposable from the standard datacube lattice. The following lemmas and propositions give the fundamental properties of the cube lattice, which are repeated in Theorem \ref{thm:treillis_cube_hierarchique}.

\begin{lemma}\label{lem:treillis_cube_hierarchique}
	The ordered set $CL(r) = \langle Space (r), \preceq_s \rangle$ is a complete lattice called cube lattice for which:
	\begin{itemize}
		\item $\forall T \subseteq CL(r), \bigwedge T = +_{t \in T}~t $ où $\bigwedge$ symbolises the \emph{infimum}. 
		\item $\forall T \subseteq CL(r), \bigvee T = \bullet_{t \in T}~t $ où $\bigvee$ symbolises the \emph{supremum}.
	\end{itemize}
\end{lemma}

\begin{lemma}\label{lem:treillis_cube_treillis_cube_hierarchique_atomique}\index{Atomic Hierarchical Cube Lattice}\index{Co-atomic Hierarchical Cube Lattice}
	The lattice $CL(r) = \langle Space (r), \preceq_s \rangle$ is a co-atomic and atomic lattice.
\end{lemma}

\begin{proposition}[Hierarchical lattice of parts of binary attributes]\label{prop:treillis_hiérarchique_des_parties_des_attributs_binaires}\index{Hierarchical lattice of parts of binary attributes}\index{Hierarchical order-embedding}
	Let $\mathcal{L}(r)$ be the hierarchical lattice of parts of binary attributes of the binary relation, \emph{i.e.} the lattice $\langle \mathcal{P}(\bigcup\limits_{d_i \in \mathcal{D}} d_i.a, \forall a \in Dom(d_i)), \subseteq \rangle$. Then there exists an order-embedding $\Phi$: \\
	$CL(r) \rightarrow \mathcal{L}(r)$ 
	\[
	t \mapsto \left\lbrace 
	\begin{array}{l}
		\bigcup\limits_{d_i \in \mathcal{D}} d_i.a, \forall a \in Dom(d_i) \text{ if } t = (\emptyset, \dotsc, \emptyset) \\
		\{d_i.t[d_i] \mid \forall d_i \in Attribute(t)\} \text{ otherwise.}
	\end{array} \right.
	\]
	The rank of a tuple $t$, noted $rank(t)$\index{Tuple rank}, is the length of the smallest path (minimal number of arcs) in the cube lattice connecting it to the tuple $(ALL_i, \dotsc, ALL_j)$. We thus have: $ rank(t) = |\Phi(t)| \text{ if } t \neq (\emptyset, \dotsc, \emptyset), |\mathcal{D}|+1 \text{ otherwise}.$
\end{proposition}

\begin{proposition}[Hierarchical co-atom / atom]\index{Hierarchical co-atom}\index{Hierarchical atom}
	Hierarchical co-atoms (respectively hierarchical atoms) are the maximal tuples, namely the most specific tuples (respectively maximal tuples) of the lattice deprived of its universal majorant (respectively minorant). The hierarchical co-atoms (respectively hierarchical atoms) of the hierarchical cube lattice of a relation $r$ are noted $\mathcal{CA}t(CL(r))$ (respectively $\mathcal{A}t(CL(r))$).
\end{proposition}

\begin{lemma}\label{lem:treillis_cube_hierarchique_gradue}\index{Graded Hierarchical Cube Lattice}
	The cube lattice $CL(r)$ is graduated. If $|\mathcal{D}| \leq 2$ then $CL(r)$ is not distributive.
\end{lemma}

Lemma \ref{lem:treillis_cube_hierarchique_gradue} shows that the cube lattice is a graduated lattice. Therefore, we can apply level-wise algorithms on this search space.

\begin{theorem}[Hierarchical datacube lattice]\label{thm:treillis_cube_hierarchique}\index{Hierarchical datacube lattice}
	Let $r$ be a data warehouse composed of hierarchical dimensions and measures ($\mathcal{D} \cup \mathcal{M}$). The ordered set $CL(r) = \langle Space (r), \preceq_s \rangle$ is a complete, atomic, co-atomic and gradual hierarchical lattice called a hierarchical datacube lattice in which:
	\begin{itemize} 
		\item $\forall T \subseteq CL(r), \bigwedge T = +_{t \in T} t$
		\item $\forall T \subseteq CL(r), \bigvee T = \bullet_{t \in T} t$.
	\end{itemize}
\end{theorem}

\begin{figure}[htbp]
	\centering
	\tiny
	\resizebox{1\textwidth}{!}{
		\begin{tikzpicture}[
			line join=bevel
			]
			
			\node (bottom) at (150pt, 0pt) {$(*, *, *)$};
			\node (nbottombd) at (175pt, 15pt) {$\ldots$};
			\node (n11) at (45pt, 75pt) {$(F, *, *)$};
			\node (n12) at (90pt, 75pt) {$(I, *, *)$};
			\node (n13) at (135pt, 75pt) {$(P, *, *)$};
			
			\node (n14) at (180pt, 75pt) {$(*, 1, *)$};
			
			\node (n15) at (225pt, 75pt) {$(*, *, 7)$};
			
			\node (n16d) at (250pt, 75pt) {$\ldots$};
			\node (n11bd) at (45pt, 90pt) {$\ldots$};
			
			\node (n12b1) at (80pt, 95pt) {};
			\node (n12bd) at (90pt, 90pt) {$\ldots$};
			\node (n12b2) at (100pt, 95pt) {};
			
			\node (n13b1) at (125pt, 95pt) {};
			\node (n13bd) at (135pt, 90pt) {$\ldots$};
			\node (n13b2) at (145pt, 95pt) {};
			
			\node (n14bd) at (180pt, 90pt) {$\ldots$};
			
			\node (n15b1) at (215pt, 95pt) {};
			\node (n15bd) at (225pt, 90pt) {$\ldots$};
			\node (n15b2) at (235pt, 95pt) {};
			\node (n21id) at (0pt, 135pt) {$\ldots$};
			
			\node (n22i1) at (20pt, 130pt) {};
			\node (n22id) at (30pt, 135pt) {$\ldots$};
			\node (n22i2) at (40pt, 130pt) {};
			
			\node (n23id) at (60pt, 135pt) {$\ldots$};
			
			\node (n24id) at (90pt, 135pt) {$\ldots$};
			
			\node (n25id) at (120pt, 135pt) {$\ldots$};
			
			\node (n26id) at (150pt, 135pt) {$\ldots$};
			
			\node (n27i1) at (170pt, 130pt) {};
			\node (n27id) at (180pt, 135pt) {$\ldots$};
			\node (n27i2) at (190pt, 130pt) {};
			
			\node (n28i1) at (200pt, 130pt) {};
			\node (n28id) at (210pt, 135pt) {$\ldots$};
			\node (n28i2) at (220pt, 130pt) {};
			
			\node (n29id) at (240pt, 135pt) {$\ldots$};
			
			\node (n210id) at (270pt, 135pt) {$\ldots$};
			\node (n21) at (0pt, 150pt) {$(F, 1, *)$};
			
			\node (n22) at (30pt, 150pt) {$(P, \operatorname{3-3}, *)$};
			\node (n23) at (60pt, 150pt) {$(M, 1, *)$};
			
			\node (n24) at (89pt, 150pt) {$(P_1, 1, *)$};
			
			\node (n25) at (118pt, 150pt) {$(F, *, 1)$};
			\node (n26) at (146pt, 150pt) {$(F, *, \operatorname{1-1})$};
			
			\node (n27) at (179pt, 150pt) {$(fr, *, \operatorname{7-12})$};
			\node (n28) at (211pt, 150pt) {$(P_1, *, 1)$};
			
			\node (n29) at (240pt, 150pt) {$(*, 1, 1)$};
			\node (n210) at (270pt, 150pt) {$(*, 1, \operatorname{1-1})$};
			
			\node (n211d) at (295pt, 150pt) {$\ldots$};
			\node (n21bd) at (-1pt, 165pt) {$\ldots$};
			
			\node (n22b1) at (20pt, 170pt) {};
			\node (n22bd) at (30pt, 165pt) {$\ldots$};
			\node (n22b2) at (40pt, 170pt) {};
			
			\node (n23b1) at (50pt, 170pt) {};
			\node (n23bd) at (60pt, 165pt) {$\ldots$};
			\node (n23b2) at (70pt, 170pt) {};
			
			\node (n24bd) at (90pt, 165pt) {$\ldots$};
			
			\node (n25bd) at (125pt, 165pt) {$\ldots$};
			
			\node (n26bd) at (150pt, 165pt) {$\ldots$};
			
			\node (n27b1) at (170pt, 170pt) {};
			\node (n27bd) at (180pt, 165pt) {$\ldots$};
			\node (n27b2) at (190pt, 170pt) {};
			
			\node (n28bd) at (215pt, 165pt) {$\ldots$};
			
			\node (n29bd) at (245pt, 165pt) {$\ldots$};
			
			\node (n210bd) at (270pt, 165pt) {$\ldots$};
			\node (n31id) at (5pt, 210pt) {$\ldots$};
			
			\node (n32id) at (30pt, 210pt) {$\ldots$};
			
			\node (n33id) at (60pt, 210pt) {$\ldots$};
			
			\node (n34id) at (90pt, 210pt) {$\ldots$};
			
			\node (n35id) at (125pt, 210pt) {$\ldots$};
			
			\node (n36id) at (150pt, 210pt) {$\ldots$};
			
			\node (n37i1) at (170pt, 205pt) {};
			\node (n37id) at (180pt, 210pt) {$\ldots$};
			\node (n37i2) at (190pt, 205pt) {};
			
			\node (n38id) at (205pt, 210pt) {$\ldots$};
			
			\node (n39id) at (240pt, 210pt) {$\ldots$};
			
			\node (n310i1) at (260pt, 205pt) {};
			\node (n310id) at (270pt, 210pt) {$\ldots$};
			\node (n310i2) at (280pt, 205pt) {};
			\node (n31) at (0pt, 225pt) {$(F, 1, 1)$};
			\node (n32) at (30pt, 225pt) {$(F, 1, \operatorname{1-1})$};
			\node (n33) at (60pt, 225pt) {$(F, 1, \operatorname{1-2})$};
			\node (n34) at (90pt, 225pt) {$(F, 1, 2)$};
			
			\node (n35) at (120pt, 225pt) {$(F, \operatorname{1-1}, 1)$};
			
			\node (n36) at (148pt, 225pt) {$(I, 1, 1)$};
			
			\node (n37) at (179pt, 225pt) {$(es, \operatorname{3-3}, \operatorname{7-12})$};
			\node (n38) at (212pt, 225pt) {$(P_1, 1, 1)$};
			
			\node (n39) at (240pt, 225pt) {$(P_1, 1, 2)$};
			
			\node (n310) at (270pt, 225pt) {$(P_3, 3, 7)$};
			
			\node (n311d) at (295pt, 225pt) {$\ldots$};
			\node (n31bd) at (270pt, 240pt) {$\ldots$};
			\node (top) at (150pt, 300pt) {$(\emptyset, \emptyset, \emptyset)$};
			
			\draw [stealth-] (bottom) -- (n11.south);
			\draw [stealth-] (bottom) -- (n12.south);
			\draw [stealth-] (bottom) -- (n13.south);
			\draw [stealth-] (bottom) -- (n14.south);
			\draw [stealth-] (bottom) -- (n15.south);
			\draw [stealth-] (n11.115) -- (n21.south);
			\draw [stealth-] (n11.90) -- (n25.south);
			\draw [stealth-] (n11.65) -- (n26.south);
			
			\draw [dashed, stealth-] (n12) -- (n12b1);
			\draw [dashed, stealth-] (n12) -- (n12b2);
			
			\draw [dashed, stealth-] (n13) -- (n13b1);
			\draw [dashed, stealth-] (n13) -- (n13b2);
			
			\draw [stealth-] (n14.140) -- (n21.south);
			\draw [stealth-] (n14.115) -- (n23.south);
			\draw [stealth-] (n14.90) -- (n24.south);
			\draw [stealth-] (n14.65) -- (n29.south);
			\draw [stealth-] (n14.40) -- (n210.south);
			
			\draw [dashed, stealth-] (n15) -- (n15b1);
			\draw [dashed, stealth-] (n15) -- (n15b2);
			\draw [dashed, stealth-] (n22i1) -- (n22.south);
			\draw [dashed, stealth-] (n22i2) -- (n22.south);
			
			\draw [dashed, stealth-] (n27i1) -- (n27.south);
			\draw [dashed, stealth-] (n27i2) -- (n27.south);
			
			\draw [dashed, stealth-] (n28i1) -- (n28.south);
			\draw [dashed, stealth-] (n28i2) -- (n28.south);
			\draw [stealth-] (n21.140) -- (n31.south);
			\draw [stealth-] (n21.115) -- (n32.south);
			\draw [stealth-] (n21.90) -- (n33.south);
			\draw [stealth-] (n21.65) -- (n34.south);
			
			\draw [dashed, stealth-] (n22) -- (n22b1);
			\draw [dashed, stealth-] (n22) -- (n22b2);
			
			\draw [dashed, stealth-] (n23) -- (n23b1);
			\draw [dashed, stealth-] (n23) -- (n23b2);
			
			\draw [stealth-] (n24.115) -- (n38.south);
			\draw [stealth-] (n24.90) -- (n39.south);
			
			\draw [stealth-] (n25.115) -- (n31.south);
			\draw [stealth-] (n25.90) -- (n35.south);
			
			\draw [stealth-] (n26.90) -- (n32.south);
			
			\draw [dashed, stealth-] (n27) -- (n27b1);
			\draw [dashed, stealth-] (n27) -- (n27b2);
			
			\draw [stealth-] (n28.90) -- (n38.south);
			
			\draw [stealth-] (n29.115) -- (n31.south);
			\draw [stealth-] (n29.90) -- (n36.south);
			\draw [stealth-] (n29.65) -- (n38.south);
			
			\draw [stealth-] (n210.90) -- (n32.south);
			\draw [dashed, stealth-] (n37i1) -- (n37.south);
			\draw [dashed, stealth-] (n37i2) -- (n37.south);
			
			\draw [dashed, stealth-] (n310i1) -- (n310.south);
			\draw [dashed, stealth-] (n310i2) -- (n310.south);
			\draw [stealth-] (n31.north) -- (top);
			\draw [stealth-] (n32.north) -- (top);
			\draw [stealth-] (n33.north) -- (top);
			\draw [stealth-] (n34.north) -- (top);
			\draw [stealth-] (n35.north) -- (top);
			\draw [stealth-] (n36.north) -- (top);
			\draw [stealth-] (n37.north) -- (top);
			\draw [stealth-] (n38.north) -- (top);
			\draw [stealth-] (n39.north) -- (top);
			\draw [stealth-] (n310.north) -- (top);
		\end{tikzpicture}
	}
	\caption{Hasse diagram of the hierarchical cube lattice of \texttt{OM3}}\label{fig:treillis_cube_hierarchique}
\end{figure}
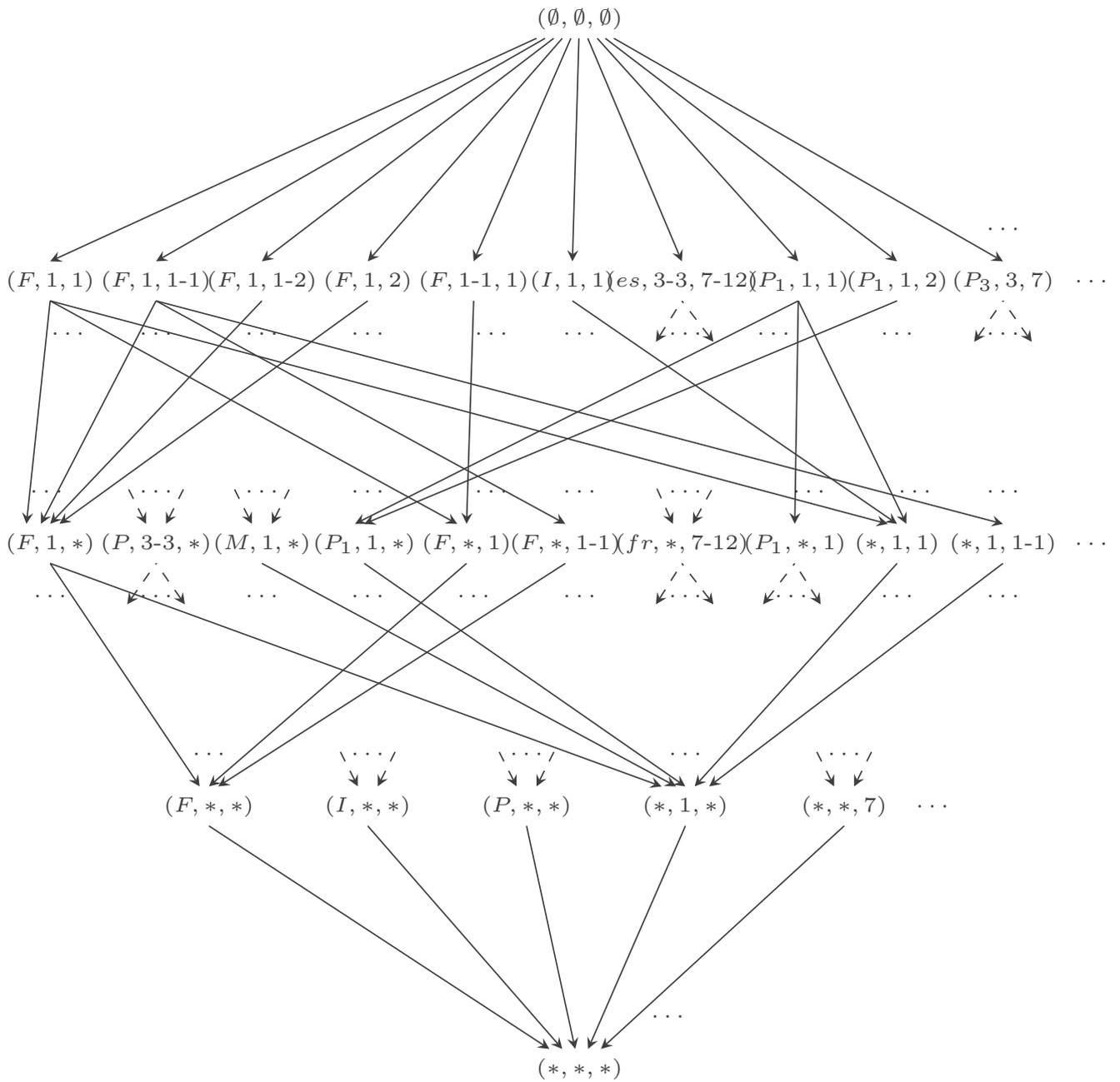

\begin{example}
	Figure \ref{fig:treillis_cube_hierarchique} shows the cube lattice of our example table of facts relation (\emph{cf.} table \ref{tab:fait_om3}). In this diagram, the edges represent the links of generalization or specialization between tuples. The values of the coded attributes are as follows:
	\begin{center}
		\begin{tabular}{ll|ll|ll}
			\toprule
			\texttt{Location} & & \texttt{Step} & & \texttt{Association} & \\
			\midrule
			$France$                & $ = F$ & $S_x (x \in \mathbb{N}^*)$        & $ = x$   & $A_x (x \in \mathbb{N}^*)$        & $ = x$   \\
			$IDF$                   & $ = I$ & $S_{x-y} (x, y \in \mathbb{N}^*)$ & $ = x-y$ & $A_{x-y} (x, y \in \mathbb{N}^*)$ & $ = x-y$ \\
			$PACA$                  & $ = P$ & $ALL_{\texttt{Turn}}$             & $ = *$   & $ALL_{\texttt{Series}}$            & $ = *$  \\
			$Marseille$             & $ = M$ &                                   &          &                                   &          \\
			$ALL_{\texttt{Player}}$ & $ = *$ &                                   &          &                                   &          \\
			\bottomrule
		\end{tabular}
	\end{center}
	
	Disregarding $(\emptyset, \emptyset, \emptyset)$, hierarchical co-atoms are tuples conveying information at the most detailed level, \emph{i.e.} that of the actual values of dimensions. In other words, the hierarchical co-atoms are the potential tuples of a relation. Thus, we have: 
	\[
	\mathcal{CA}t(CL(\texttt{OM3})) = 
	\left\lbrace
	\begin{array}{ll}
		{(France, S_1, A_1)},      \\
		{(France, S_1, A_{1-1})},  \\
		{(France, S_1, A_{1-2})},  \\
		{(France, S_1, A_2)},      \\
		\ldots                     \\
		{(France, S_{1-1}, A_1)},  \\
		\ldots                     \\
		{(IDF, S_1, A_1)},         \\
		\ldots                     \\
		{(es, S_{3-3}, A_{7-12})}, \\
		{(P_1, S_1, A_1)},         \\
		\ldots                     \\
		{(P_1, S_1, A_2)},         \\
		\ldots                     \\
		{(P_2, S_3, A_7)},         \\
		\ldots                            
	\end{array}
	\right\rbrace     
	\]
	
	The hierarchical atoms of the lattice offer the most synthetic information possible, with the exception of the tuple ($ALL_{\texttt{Player}}$, $ALL_{\texttt{Turn}}$, $ALL_{\texttt{Series}}$). Since we will only consider three dimensions here, all hierarchical atoms have two different synthetic $ALL_i$ values. Thus we have:
	\[\mathcal{A}t(CL(r)) = 
	\left\lbrace
	\begin{array}{ll}
		{(France, ALL_{\texttt{Turn}}, ALL_{\texttt{Series}})}, \\
		{(IDF, ALL_{\texttt{Turn}}, ALL_{\texttt{Series}})},    \\
		{(Paris, ALL_{\texttt{Turn}}, ALL_{\texttt{Series}})},  \\
		\ldots                                                  \\
		{(ALL_{\texttt{Player}}, S_1, ALL_{\texttt{Series}})},  \\
		\ldots                                                  \\
		{(ALL_{\texttt{Player}}, ALL_{\texttt{Turn}}, A_7)},    \\
		\ldots                                                        
	\end{array}
	\right\rbrace
	\]
\end{example}

\begin{property}[Size of a hierarchical cube lattice]\index{Size of a hierarchical cube lattice}
	The height (number of levels) of the hierarchical cube lattice is $|\mathcal{D}| + 1$. The number of elements for a level $i (i \in 1..|\mathcal{D}|)$ is:
	\[\sum\limits_{\scriptstyle d_i \subseteq \mathcal{D} \atop \scriptstyle |d_i| = i} (\prod\limits_{A \in d_i} |Dom(h_i)|) \leq \binom{|\mathcal{D}|}{i} \max_{A \in \mathcal{D}}(|Dom(h_i)|)^i.\]
	
	The total number of elements in the hierarchical cube lattice is:
	\[\sum_{i=1..|\mathcal{D}|}(\sum\limits_{\scriptstyle d_i \subseteq \mathcal{D} \atop \scriptstyle |d_i| = i} (\prod\limits_{A \in d_i} |Dom(h_i)|)) + 2 = (\prod\limits_{A \in \mathcal{D}} (|Dom(h_i))| + 1)) + 1\]
\end{property}

The above property gives an analysis of the number of elements contained in a level of the hierarchical cube lattice and the total number of elements. This property is particularly important as it allows us to characterise the complexity of algorithms using the hierarchical cube lattice as a search space.

\section{Hierarchical datacube}

The concept of a datacube in a relational world (\cite{grayDataCubeRelational1997}) is immediately transposed into a hierarchical datacube in a data warehouse world, with the integration of hierarchical dimensions.

Thus, we call the result of a partitioning query, according to a set $X \subseteq \mathcal{D}$ of hierarchical dimensions, cuboid, noted $cuboid_{X}(r)$. Likewise, the datacube constituted by the set of all cuboids is noted $datacube(r)$. The cuboids can be ordered with respect to their level of detail by the partial order relation $\preceq$.

\begin{example}
	The cuboid according to the hierarchical dimensions $\texttt{D}_{\texttt{Player}}, \texttt{D}_{\texttt{Turn}}$ is less aggregated (more detailed) than the cuboid according to $\texttt{D}_{\texttt{Player}}$ or that according to $\texttt{D}_{\texttt{Turn}}$. In other words:
	\begin{align*}
		&Cuboid_{\texttt{D}_{\texttt{Player}}}(r) \preceq Cuboid_{\texttt{D}_{\texttt{Player}},\texttt{D}_{\texttt{Turn}}}(r) \\
		&Cuboid_{\texttt{D}_{\texttt{Turn}}}(r) \preceq Cuboid_{\texttt{D}_{\texttt{Player}},\texttt{D}_{\texttt{Turn}}}(r)
	\end{align*}
\end{example}

In a standard way, the set of cuboids endowed with this order forms a lattice. The cuboids of this lattice are grouped by level according to their number of hierarchical dimensions. These levels are numbered starting from the bottom of the lattice (cuboid bearing no dimension $d_i$) and going up towards the top (cuboid according to all possible criteria called "base cuboid"). Let's consider two cuboids $cuboid_{\mathcal{U}}(r)$ according to the hierarchical dimension subset $\mathcal{U}$ and $cuboid_{\mathcal{V}}(r)$ according to $\mathcal{V}$, if $\mathcal{U} \subset \mathcal{V}$ then $cuboid_{\mathcal{U}}(r) \preceq cuboid_{\mathcal{V}}(r)$, the cuboids are said to have a family relationship, $cuboid_{\mathcal{V}}(r)$ is called the "ancestor" of $cuboid_{\mathcal{U}}(r)$ and $cuboid_{\mathcal{U}}(r)$ is the "offspring" of $cuboid_{\mathcal{V}}(r)$.

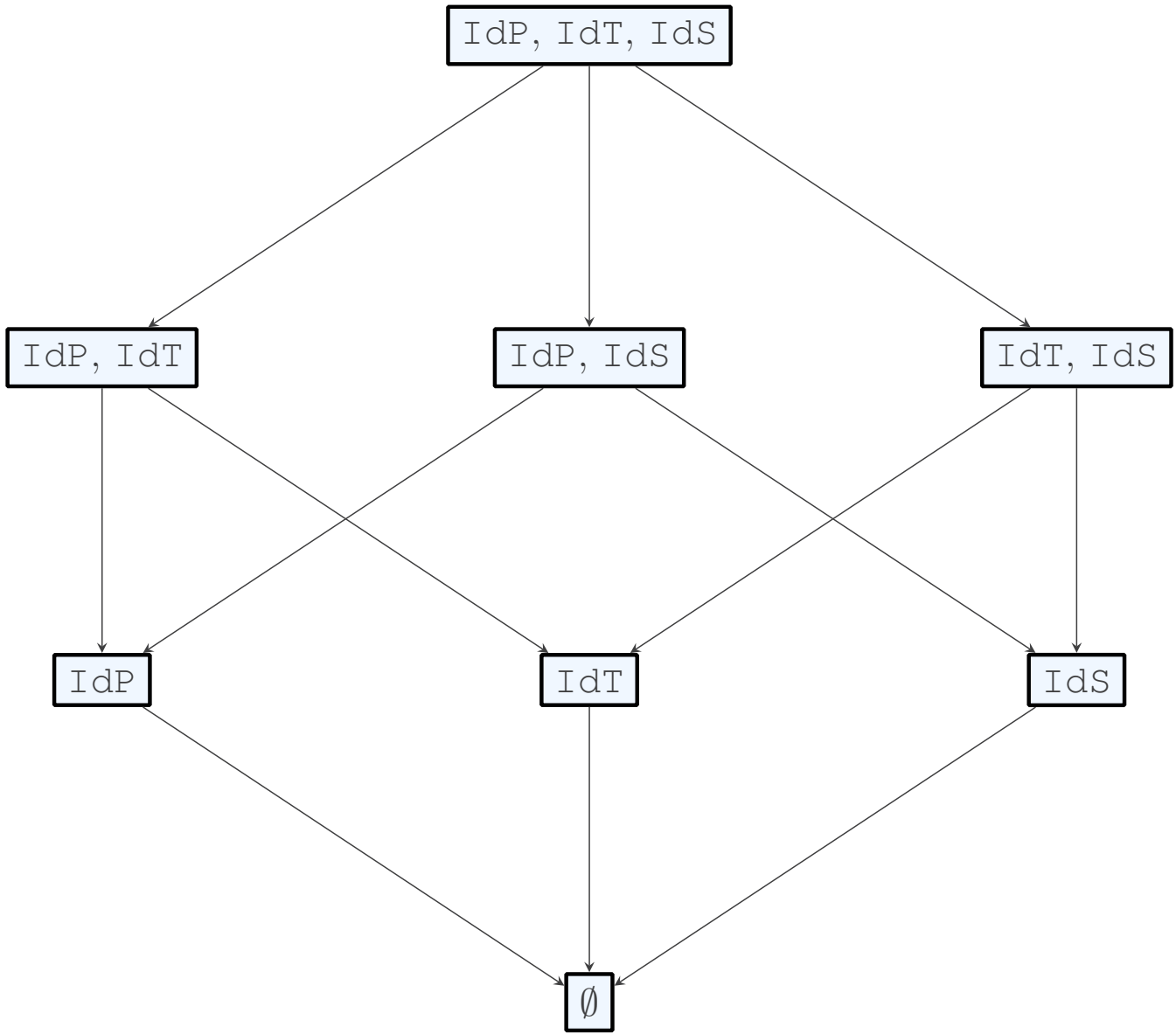
\begin{figure}[htbp]
	\centering
	\large
	\resizebox{1\textwidth}{!}{
		\begin{tikzpicture}[
			line join=bevel,
			box/.style={rectangle, draw=black, fill=aliceblue, very thick, minimum size=5mm}
			]
			
			\node [box] (bottom) at (150pt, 0pt) {$\emptyset$};
			\node [box] (n11) at (0pt, 100pt) {\texttt{IdP}};
			\node [box] (n12) at (150pt, 100pt) {\texttt{IdT}};
			\node [box] (n13) at (300pt, 100pt) {\texttt{IdS}};
			\node [box] (n21) at (0pt, 200pt) {\texttt{IdP}, \texttt{IdT}};
			\node [box] (n22) at (150pt, 200pt) {\texttt{IdP}, \texttt{IdS}};
			\node [box] (n23) at (300pt, 200pt) {\texttt{IdT}, \texttt{IdS}};
			\node [box] (top) at (150pt, 300pt) {\texttt{IdP}, \texttt{IdT}, \texttt{IdS}};
			
			\draw [stealth-] (bottom) -- (n11);
			\draw [stealth-] (bottom) -- (n12);
			\draw [stealth-] (bottom) -- (n13);
			\draw [stealth-] (n11) -- (n21);
			\draw [stealth-] (n11) -- (n22);
			\draw [stealth-] (n12) -- (n21);
			\draw [stealth-] (n12) -- (n23);
			\draw [stealth-] (n13) -- (n22);
			\draw [stealth-] (n13) -- (n23);
			\draw [stealth-] (n21) -- (top);
			\draw [stealth-] (n22) -- (top);
			\draw [stealth-] (n23) -- (top);
		\end{tikzpicture}
	}
	\caption{Cuboids of \textsc{OM3} according to $\texttt{D}_{\texttt{Player}}$ (\texttt{IdP}), $\texttt{D}_{\texttt{Turn}}$ (\texttt{IdT}) and $\texttt{D}_{\texttt{Series}}$ (\texttt{IdS})}\label{fig:cuboid_treillis_dimensions_hierachiques}
\end{figure}

Figure \ref{fig:cuboid_treillis_dimensions_hierachiques} gives the representation of the cuboid lattice of the data warehouse \texttt{OM3} according to the hierarchical dimensions $\texttt{D}_{\texttt{Player}}$ (\texttt{IdP}), $\texttt{D}_{\texttt{Turn}}$ (\texttt{IdT}) and $\texttt{D}_{\texttt{Series}}$ (\texttt{IdS}).

By taking the example of the \texttt{OM3} and its table of facts (\emph{cf.} table \ref{tab:fait_om3}), we can express in SQL the hierarchical datacube necessary for an analysis according to the measurements \texttt{Time} and \texttt{Score} with respect to hierarchical dimensions \texttt{IdP}, \texttt{IdT} and \texttt{IdS} with the following query:

\lstset{style=SQLStyle}
\begin{lstlisting}[ language=SQL,
					deletekeywords={IDENTITY},
					deletekeywords={[2]INT},
					morekeywords={CLUSTERED, SKYLINE, OF},
					framesep=8pt,
					xleftmargin=40pt,
					framexleftmargin=40pt,
					frame=tb,
					framerule=0pt ]
	SELECT IdP, IdT, IdS, SUM(Time), MAX(Score)
	FROM OM3
	GROUP BY CUBE IdP, IdT, IdS
\end{lstlisting}

This query will calculate the $2^3 = 8$ partitionings (where $\emptyset$ represents the partitioning along no dimension $d_i$, \emph{i.e.} aggregating the entire data warehouse into a single tuple:
\begin{itemize}
	\item $'\texttt{IdP}, \texttt{IdT}, \texttt{IdS}'$~;
	\item $'\texttt{IdP}, \texttt{IdT}'$~;
	\item $'\texttt{IdP}, \texttt{IdS}'$~;
	\item $'\texttt{IdT}, \texttt{IdS}'$~;
	\item $'\texttt{IdP}'$~;
	\item $'\texttt{IdT}'$~;
	\item $'\texttt{IdS}'$~;
	\item $'\emptyset'$.
\end{itemize}

As with the traditional datacube, the naive way to compute this type of query is to rewrite it as a collection of eight aggregative queries and run them separately. However, each of its subqueries has its own pattern. To make all these cuboids uni-compatible, we use the value $ALL_i$ (\emph{cf.} definition \ref{def:element_maximal_d_une_hierarchie}): this value has a particular semantics, it is a generalization of all the values of the domain of an attribute of a dimension $d_i$. Thus, the cuboids all share the same schema ($\mathcal{D} \cup \mathcal{M}$) and can be grouped together in a single relation: the hierarchical datacube. Each multidimensional tuple consists of a set of values for its hierarchical dimensions and numerical values for their measures. The value of the measure is computed by aggregating the set of tuples of the original relation sharing the same values of the selected hierarchical dimensions.

\begin{example}
	The multidimensional tuple $t = (8, 1, ALL)$ is obtained by aggregating by \texttt{IdP} and \texttt{IdT} the tuples $t_1 = (8, 1, 1)$ and $t_2 = (8, 1, 4)$ of the example data warehouse (\emph{cf.} table \ref{tab:fait_om3}). The measures of $t$ are obtained by applying the measure functions on the set of tuples needed for the aggregation. 
	
	In our case, we have:
	\begin{itemize}
		\item $f_{\texttt{Time}}(t,r) = t_1(\texttt{Time})  + t_2(\texttt{Time}) = SUM(6.32, 18.9) = 25.22$
		\item \ldots
		\item $f_{\texttt{Score}}(t,r) = t_1(\texttt{Score}) +t_2(\texttt{Score}) = MAX(700, 2300) = 2300$
		\item \ldots
	\end{itemize}
\end{example}		

With the example data warehouse (\emph{cf.} table \ref{tab:fait_om3}), the query to naively compute the hierarchical datacube according to the hierarchical dimensions $\texttt{D}_{\texttt{Player}}$ (\texttt{IdP}), $\texttt{D}_{\texttt{Turn}}$ (\texttt{IdT}) and $\texttt{D}_{\texttt{Series}}$ (\texttt{IdS}) and the measures \texttt{Time}, \texttt{Duration}, \texttt{Number}, \texttt{Score} and \texttt{Shape} is as follows:

\lstset{style=SQLStyle}
\begin{lstlisting}[ language=SQL,
					deletekeywords={IDENTITY},
					deletekeywords={[2]INT},
					morekeywords={CLUSTERED, SKYLINE, OF},
					framesep=8pt,
					xleftmargin=40pt,
					framexleftmargin=40pt,
					frame=tb,
					framerule=0pt ]
	SELECT IdP, IdT, IdS, SUM(Time), MAX(Score)
	FROM OM3
	GROUP BY IdP, IdT, IdS
	UNION
	SELECT IdP, IdT, ALL, SUM(Time), MAX(Score)
	FROM OM3
	GROUP BY IdP, IdT
	UNION
	SELECT IdP, ALL, IdS, SUM(Time), MAX(Score)
	FROM OM3
	GROUP BY IdP, IdS
	UNION
	SELECT ALL, IdT, IdS, SUM(Time), MAX(Score)
	FROM OM3
	GROUP BY IdT, IdS
	UNION
	SELECT IdP, ALL, ALL, SUM(Time), MAX(Score)
	FROM OM3
	GROUP BY IdP
	UNION
	SELECT ALL, IdT, ALL, SUM(Time), MAX(Score)
	FROM OM3
	GROUP BY IdT
	UNION
	SELECT ALL, ALL, IdS, SUM(Time), MAX(Score)
	FROM OM3
	GROUP BY IdS
	UNION
	SELECT ALL, ALL, ALL, SUM(Time), MAX(Score)
	FROM OM3
\end{lstlisting}

The result of this query is given by the table \ref{tab:cube_de_donnes_hierarchique}. For clarity, different cuboids are separated by a horizontal line, and special values are abbreviated: $ALL_{\texttt{Player}}$ as $ALL_{\texttt{P}}$, $ALL_{\texttt{Turn}}$ as $ALL_{\texttt{T}}$ and $ALL_{\texttt{Series}}$ as $ALL_{\texttt{S}}$.

\begin{table}[htbp]
	\caption{Relational representation of the hierarchical datacube of \texttt{OM3}}\label{tab:cube_de_donnes_hierarchique}
	\centering
	\begin{tabular}{ccc|ccccc}
		\toprule
		\texttt{IdP} & \texttt{IdT} & \texttt{IdS} & \texttt{Time} & \texttt{Duration} & \texttt{Number} & \texttt{Score} & \texttt{Shape} \\
		\midrule
		$P_1$              & $S_1$              & $A_1$              & $6.32$   & $2.85$ & $3.5$  & $700$  & $0.5$  \\
		$P_1$              & $S_1$              & $A_2$              & $18.9$   & $1.95$ & $3.83$ & $2300$ & $0.5$  \\
		$P_2$              & $S_2$              & $A_3$              & $26.39$  & $1.7$  & $3.43$ & $2400$ & $0.71$ \\
		$P_2$              & $S_2$              & $A_4$              & $4.1$    & $2.07$ & $3$    & $300$  & $0.5$  \\
		$P_2$              & $S_2$              & $A_5$              & $7.38$   & $3.68$ & $3$    & $600$  & $0.5$  \\
		$P_3$              & $S_3$              & $A_6$              & $2.14$   & $2.15$ & $3$    & $300$  & $1$    \\
		$P_3$              & $S_3$              & $A_7$              & $56.04$  & $2.25$ & $3.17$ & $3800$ & $0.5$  \\
		\hline
		$P_1$              & $S_1$              & $ALL_{\texttt{S}}$ & $25.22$  & $2.04$ & $3.8$  & $2300$ & $0.5$  \\
		$P_2$              & $S_2$              & $ALL_{\texttt{S}}$ & $37.87$  & $1.79$ & $3.4$  & $2400$ & $0.7$  \\
		$P_3$              & $S_3$              & $ALL_{\texttt{S}}$ & $58.18$  & $2.25$ & $3.17$ & $3800$ & $0.5$  \\
		\hline
		$P_1$              & $ALL_{\texttt{T}}$ & $A_1$              & $6.32$   & $2.85$ & $3.5$  & $700$  & $0.5$  \\
		$P_1$              & $ALL_{\texttt{T}}$ & $A_2$              & $18.9$   & $1.95$ & $3.83$ & $2300$ & $0.5$  \\
		$P_2$              & $ALL_{\texttt{T}}$ & $A_3$              & $26.39$  & $1.7$  & $3.43$ & $2400$ & $0.71$ \\
		$P_2$              & $ALL_{\texttt{T}}$ & $A_4$              & $4.1$    & $2.07$ & $3$    & $300$  & $0.5$  \\
		$P_2$              & $ALL_{\texttt{T}}$ & $A_5$              & $7.38$   & $3.68$ & $3$    & $600$  & $0.5$  \\
		$P_3$              & $ALL_{\texttt{T}}$ & $A_6$              & $2.14$   & $2.15$ & $3$    & $300$  & $1$    \\
		$P_3$              & $ALL_{\texttt{T}}$ & $A_7$              & $56.04$  & $2.25$ & $3.17$ & $3800$ & $0.5$  \\
		\hline
		$ALL_{\texttt{P}}$ & $S_1$              & $A_1$              & $6.32$   & $2.85$ & $3.5$  & $700$  & $0.5$  \\
		$ALL_{\texttt{P}}$ & $S_1$              & $A_2$              & $18.9$   & $1.95$ & $3.83$ & $2300$ & $0.5$  \\
		$ALL_{\texttt{P}}$ & $S_2$              & $A_3$              & $26.39$  & $1.7$  & $3.43$ & $2400$ & $0.71$ \\
		$ALL_{\texttt{P}}$ & $S_2$              & $A_4$              & $4.1$    & $2.07$ & $3$    & $300$  & $0.5$  \\
		$ALL_{\texttt{P}}$ & $S_2$              & $A_5$              & $7.38$   & $3.68$ & $3$    & $600$  & $0.5$  \\
		$ALL_{\texttt{P}}$ & $S_3$              & $A_6$              & $2.14$   & $2.15$ & $3$    & $300$  & $1$    \\
		$ALL_{\texttt{P}}$ & $S_3$              & $A_7$              & $56.04$  & $2.25$ & $3.17$ & $3800$ & $0.5$  \\
		\hline
		$P_1$              & $ALL_{\texttt{T}}$ & $ALL_{\texttt{S}}$ & $25.22$  & $2.04$ & $3.8$  & $2300$ & $0.5$  \\
		$P_2$              & $ALL_{\texttt{T}}$ & $ALL_{\texttt{S}}$ & $37.87$  & $1.79$ & $3.4$  & $2400$ & $0.7$  \\
		$P_3$              & $ALL_{\texttt{T}}$ & $ALL_{\texttt{S}}$ & $58.18$  & $2.25$ & $3.17$ & $3800$ & $0.5$  \\
		\hline
		$ALL_{\texttt{P}}$ & $S_1$              & $ALL_{\texttt{S}}$ & $25.22$  & $2.04$ & $3.8$  & $2300$ & $0.5$  \\
		$ALL_{\texttt{P}}$ & $S_2$              & $ALL_{\texttt{S}}$ & $37.87$  & $1.79$ & $3.4$  & $2400$ & $0.7$  \\
		$ALL_{\texttt{P}}$ & $S_3$              & $ALL_{\texttt{S}}$ & $58.18$  & $2.25$ & $3.17$ & $3800$ & $0.5$  \\
		\hline
		$ALL_{\texttt{P}}$ & $ALL_{\texttt{T}}$ & $A_1$              & $6.32$   & $2.85$ & $3.5$  & $700$  & $0.5$  \\
		$ALL_{\texttt{P}}$ & $ALL_{\texttt{T}}$ & $A_2$              & $18.9$   & $1.95$ & $3.83$ & $2300$ & $0.5$  \\
		$ALL_{\texttt{P}}$ & $ALL_{\texttt{T}}$ & $A_3$              & $26.39$  & $1.7$  & $3.43$ & $2400$ & $0.71$ \\
		$ALL_{\texttt{P}}$ & $ALL_{\texttt{T}}$ & $A_4$              & $4.1$    & $2.07$ & $3$    & $300$  & $0.5$  \\
		$ALL_{\texttt{P}}$ & $ALL_{\texttt{T}}$ & $A_5$              & $7.38$   & $3.68$ & $3$    & $600$  & $0.5$  \\
		$ALL_{\texttt{P}}$ & $ALL_{\texttt{T}}$ & $A_6$              & $2.14$   & $2.15$ & $3$    & $300$  & $1$    \\
		$ALL_{\texttt{P}}$ & $ALL_{\texttt{T}}$ & $A_7$              & $56.04$  & $2.25$ & $3.17$ & $3800$ & $0.5$  \\
		\hline
		$ALL_{\texttt{P}}$ & $ALL_{\texttt{T}}$ & $ALL_{\texttt{S}}$ & $121.27$ & $2.11$ & $3.32$ & $3800$ & $0.55$ \\
		\bottomrule
	\end{tabular}
\end{table}

\section{Closed hierarchical datacube}

The combinatorial explosion of results during datacube computations is a well known phenomenon (\cite{hanDataMiningConcepts2011}), which is amplified with the computation of hierarchical datacubes. The closed hierarchical cube approach we propose aims at representing the hierarchical datacube without loss of information but with a consequent decrease of the necessary storage space. To do this, we eliminate possible redundancies by keeping only one representative for a set of tuples from the same data of the original relation. It is easily transposable from the standard closed datacube.

\subsection{Experimental results for closed datacube}

Our objective now is to compare, through various experiments, the sizes of the datacube and the closed cube. In this sub-section, we report a summary of our results. All experiments are conducted on an Intel Pentium G2120 3.10~GHz with 3.6~GB main memory and running on Turnkey LAMP Stack (based on Debian GNU/Linux). We use the algorithm Close \cite{pasquierEfficientMiningAssociation1999} (for which we have the sources) in order to perform experimental comparisons between the representations. We use real data sets to evaluate the effectiveness of our approach.

We use the real dataset \texttt{SEP85L} containing weather conditions at various weather stations from December 1981 through November 1991. This weather dataset has been frequently used in calibrating various cube algorithms \cite{weiCondensedCubeEfficient2002}. \texttt{Mushroom} is a dataset widely known in frequent pattern mining. It provides various characteristics of mushrooms. \texttt{Death} is a dataset gathering information about patients' decease with the date and cause. \texttt{TombNecropolis} and \texttt{TombObjects} are issued from archaeological excavation. They encompass a list of necropolises, their tombs and other properties like the country, the funeral rite, the objects discovered in the tombs and their description. Finally, \texttt{Joint\_Objects\_Tombs} results from the natural join between \texttt{TombObjects} and Tomb\-Ne\-cro\-po\-lis according to the identifiers of necropolises and tombs.

Table \ref{tab:desc_dataset} gives the datasets used for experiments. The columns \texttt{Att\-ributes} and \texttt{Tuples} stand for the number of attributes and tuples respectively. In the last column, the size in bytes of the dataset is reported (each dimension or attribute is encoded as an integer requiring 4 bytes for any value).

\begin{table}[htbp]
	\centering
	\caption{Experimental datasets}\label{tab:desc_dataset}
	\begin{tabular}{l|c|r|r}
		\toprule
		Tables & \texttt{Attributes} & \texttt{Tuples} & Size \\
		\midrule
		\texttt{SEP85L}                & 20 & 507\,684 & 56\,520  \\
		\texttt{Mushroom}              & 23 & 8\,124   & 747\,408 \\
		\texttt{Death}                 &  5 & 389      &   7\,780 \\
		\midrule
		\texttt{TombNecropolis}        & 7  & 1\,846   &  51\,688 \\
		\texttt{TombObjects}           & 12 & 8\,278   & 397\,344 \\
		\texttt{Joint\_Objects\_Tombs} & 17 & 7\,643   & 519\,724 \\
		\bottomrule
	\end{tabular}  
\end{table}

Table \ref{table::tailleCube} illustrates the size of the studied representations for the various datasets.

\begin{table}[htbp]
	\centering
	\caption{Size of the Data Cubes (in bytes)}\label{table::tailleCube}
	\begin{tabular}{l|r|r}
		\toprule
		& Datacube & Closed cube \\
		\midrule
		\texttt{Death}                        & 220\,152      & 24\,984     \\
		\texttt{TombNecropolis}               & 3\,639\,072   & 189\,728    \\
		\texttt{Joint\_Objects\_Tombs} (1~\%) & 58\,848\,264  & 4\,485\,168 \\
		\texttt{Mushroom} (5~\%)              & 436\,823\,808 & 1\,233\,984 \\
		\texttt{TombObjects}                  & 903\,611\,124 & 8\,032\,648 \\
		\bottomrule 
	\end{tabular}
\end{table}

These five datasets are only encompassing strongly correlated data. Thus we are in the most difficult cases. In this context, the closed cube reduces the size of the datacube from 8 to more than 300 times. 

By using the \texttt{SEP85L} dataset, we have generated 9 datasets having from 2 to 10 dimensions by projecting the weather dataset on the first $k$ dimensions (2 $\geq\ k \geq 10$). Table \ref{tab:eva_expe1} presents the number of resulting patterns for the  approaches.

\begin{table}[htbp]
	\centering
	\caption{Experimental results for \texttt{SEP85L}}\label{tab:eva_expe1}
	\begin{tabular}{l|r|r}
		\toprule
		& \multicolumn{2}{c}{Number of tuples} \\
		\midrule
		Dimensions & Datacube & Closed datacube \\
		\midrule
		2  & 12\,234  & 7\,928   \\
		3  & 21\,114  & 13\,883  \\
		4  & 27\,514  & 18\,874  \\
		5  & 38\,793  & 24\,478  \\
		6  & 62\,248  & 34\,898  \\
		7  & 109\,314 & 52\,344  \\
		8  & 199\,639 & 81\,640  \\
		9  & 371\,791 & 136\,275 \\
		10 & 722\,133 & 172\,452 \\
		\bottomrule
	\end{tabular}
\end{table}

Whatever the number of dimensions is, the closed cube is the smallest representation. It is always significantly reduced when compared to the datacube itself. 

\subsection{Formal framework of closed hierarchical datacube}

The following definitions and corollary give the fundamental properties of the closed hierarchical datacube, which are repeated in theorem \ref{thm:treillis_cube_hierarchique_ferme}.

\color{black}

\begin{definition}[Hierarchical cube closure operator]\label{def:operateur_de_fermeture_cubique_hierarchique}\index{Hierarchical cube closure operator}\index{Hierarchical cube closure}
	The hierarchical cube closure operator associates to any tuple $t$ of the hierarchical cube lattice a single tuple called the hierarchical closure of $t$, or simply the closure of $t$. It is obtained by considering the set of tuples more specific than $t$ and by determining the most general tuple of this set using the Sum operator. This operator is noted $\mathbb{C}$ and defined as follows:
	\[
	\begin{array}{lcl}
		\mathbb{C} &:& CL(r) \rightarrow CL(r)\\
		t &\mapsto&\left\lbrace
		\begin{array}{l}
			+ t' \mid t \preceq_{s} t' \text{ and } t' \in r \\
			(\emptyset, \dotsc, \emptyset)\  \text{ otherwise.}
		\end{array} \right.
	\end{array} 
	\]
\end{definition}

\begin{example}\label{ex:operateur_de_fermeture_hierarchique}
	Considering the multidimensional space of the data warehouse \texttt{OM3} (\emph{cf.} table \ref{tab:espace_multidimensionnel_hierarchique}) and its table of facts (\emph{cf.} table \ref{tab:fait_om3}), we have $\mathbb{C}((P_1, ALL_\texttt{T}, ALL_\texttt{S})) = (P_1, S_1, A_1) + (P_1, S_1, A_2) = (P_1, S_1, ALL_\texttt{S})$ and $\mathbb{C}((ALL_\texttt{P}, ALL_\texttt{T}, A_7)) = (P_3, S_3, A_7)$.
\end{example}

\begin{corollaire}
	$\mathbb{C}$ is a closure operator of $CL(r)$ on $r$. $\mathbb{C}$ satisfies the following properties:
	\begin{itemize}
		\item $t \preceq_{g} t' \Rightarrow \mathbb{C}(t, r$) $\preceq_{g}$
		$\mathbb{C}(t', r)$ (isotonicity);
		\item  $t \preceq_{g} \mathbb{C}(t,r)$ (extensivity);
		\item $\mathbb{C}(t, r) = \mathbb{C}(\mathbb{C}(t, r), r)$ (idempotency).
	\end{itemize}
\end{corollaire}

\begin{definition}[Hierarchical cube closure system]\label{def:systeme_de_fermeture_cubique_hierarchique}\index{Hierarchical cube closure system}
	Let's consider $\mathbb{C}(r)$ = \{ $t \in CL(r) \mid \mathbb{C}(t, r) = t$\}. $\mathbb{C}(r)$ is a closure system on $r$ and the associated closure operator is $\mathbb{C}$. Any tuple belonging to $\mathbb{C}(r)$ is a hierarchical closed tuple or a hierarchical cube closure.
\end{definition}

\begin{example}
	Considering the multidimensional space of the data warehouse \texttt{OM3} (\emph{cf.} table \ref{tab:espace_multidimensionnel_hierarchique}) and its table of facts (\emph{cf.} table \ref{tab:fait_om3}, we have:
	\[\mathbb{C}(r) =
	\left\lbrace
	\begin{array}{l}
		(P_1, S_1, A_1),               \\
		(P_1, S_1, A_2),               \\
		(P_2, S_2, A_3),               \\
		(P_2, S_2, A_4),               \\
		(P_2, S_2, A_5),               \\
		(P_3, S_3, A_6),               \\
		(P_3, S_3, A_7),               \\
		(P_1, S_1, ALL_\texttt{S}),    \\
		(P_2, S_2, ALL_\texttt{S}),    \\
		(P_3, S_3, ALL_\texttt{S}),    \\
		(\emptyset, \dotsc, \emptyset) \\
	\end{array}
	\right\rbrace     
	\]
\end{example}

\begin{theorem}\label{thm:treillis_cube_hierarchique_ferme}\index{Closed hierarchical cube lattice}
	The partially ordered set $CCL(r) = \langle \mathbb{C}(r), \preceq_g \rangle$ is a complete and co-atomic lattice called a closed cube lattice such that:
	\begin{itemize}
		\item $\forall T \subseteq CCL(r), \bigwedge T = +_{t \in T} t$
		\item $\forall T \subseteq CCL(r), \bigvee T = \mathbb{C}(\bullet_{t \in T} t,r)$
	\end{itemize}
\end{theorem}

All tuples with the same closure generalize the same tuples of the original relation. As they generalize the same original tuples, they share the same aggregated value of the measure (property of \texttt{GROUP BY}). For each tuple of the cube lattice, it is sufficient to calculate its closure to find its measure. Thus the closed hierarchical cube is a cover of the hierarchical datacube.   

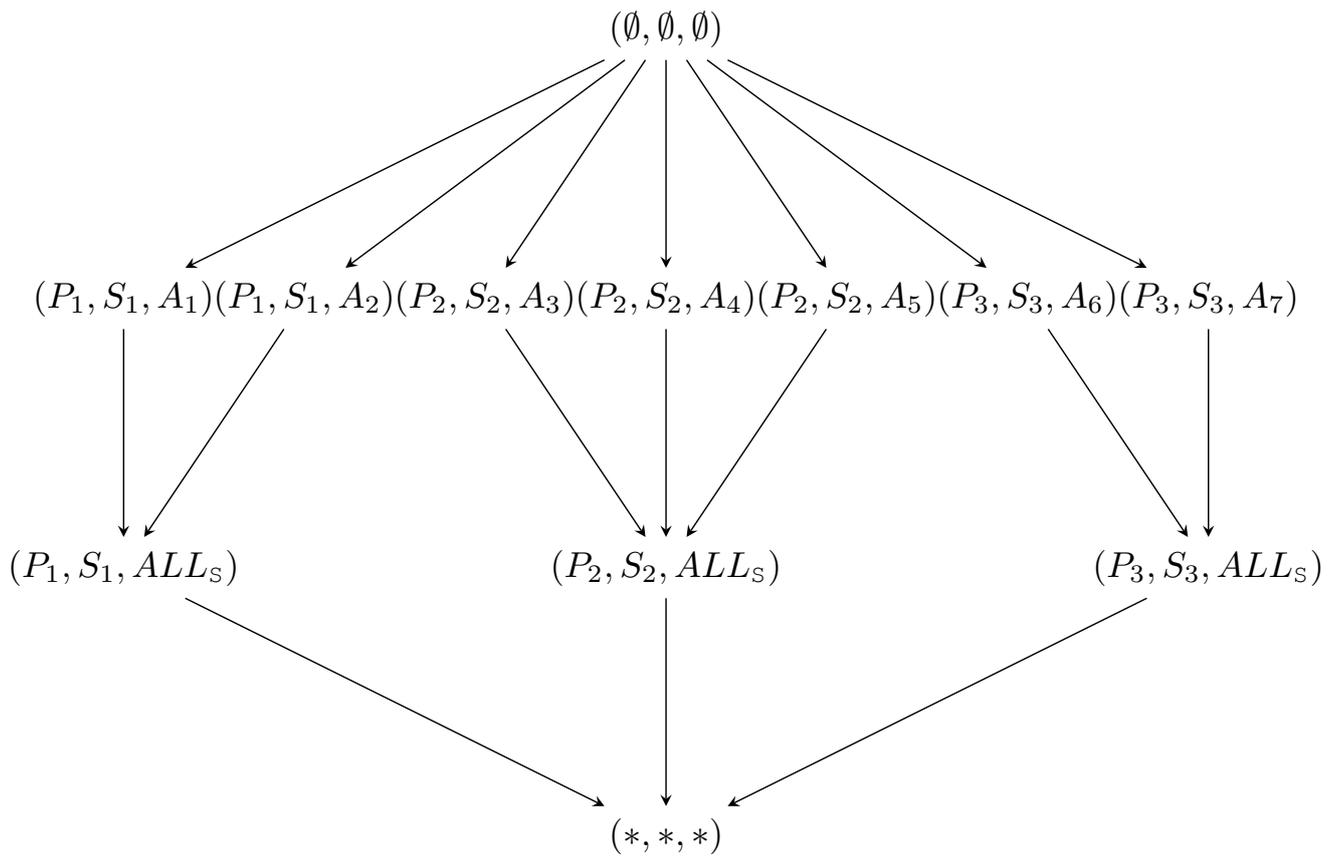
\begin{figure}[htbp]
	\centering
	\small
	\resizebox{1\textwidth}{!}{
		\begin{tikzpicture}[
			line join=bevel
			]
			
			\node (all) at (150pt, 0pt) {$(*, *, *)$};
			\node (n11) at (0pt, 75pt) {$(P_1, S_1, ALL_\texttt{S})$};
			\node (n12) at (150pt, 75pt) {$(P_2, S_2, ALL_\texttt{S})$};
			\node (n13) at (300pt, 75pt) {$(P_3, S_3, ALL_\texttt{S})$};
			\node (n21) at (0pt, 150pt) {$(P_1, S_1, A_1)$};
			\node (n22) at (50pt, 150pt) {$(P_1, S_1, A_2)$};
			\node (n23) at (100pt, 150pt) {$(P_2, S_2, A_3)$};
			\node (n24) at (150pt, 150pt) {$(P_2, S_2, A_4)$};
			\node (n25) at (200pt, 150pt) {$(P_2, S_2, A_5)$};
			\node (n26) at (250pt, 150pt) {$(P_3, S_3, A_6)$};
			\node (n27) at (300pt, 150pt) {$(P_3, S_3, A_7)$};
			\node (vide) at (150pt, 225pt) {$(\emptyset, \emptyset, \emptyset)$};
			
			\draw [stealth-] (all) -- (n11);
			\draw [stealth-] (all) -- (n12);
			\draw [stealth-] (all) -- (n13);			
			\draw [stealth-] (n11) -- (n21);
			\draw [stealth-] (n11) -- (n22);
			
			\draw [stealth-] (n12) -- (n23);
			\draw [stealth-] (n12) -- (n24);
			\draw [stealth-] (n12) -- (n25);
			
			\draw [stealth-] (n13) -- (n26);
			\draw [stealth-] (n13) -- (n27);
			\draw [stealth-] (n21) -- (vide);
			\draw [stealth-] (n22) -- (vide);
			\draw [stealth-] (n23) -- (vide);
			\draw [stealth-] (n24) -- (vide);
			\draw [stealth-] (n25) -- (vide);
			\draw [stealth-] (n26) -- (vide);
			\draw [stealth-] (n27) -- (vide);
		\end{tikzpicture}
	}
	\caption{Hasse diagram of the closed hierarchical cube lattice of \texttt{OM3}}\label{fig:treillis_cube_hierarchique_ferme}
\end{figure}

\begin{example}
	Figure \ref{fig:treillis_cube_hierarchique_ferme} shows the closed hierarchical cube lattice of \texttt{OM3}.
\end{example}

\subsection{Possible limitations of closed hierarchical datacube}

Despite offering advantages in term of storage efficiency and query acceleration, closed hierarchical datacube have some possible limitations and limitations to account for, like loss in details, analysis flexibility, hierarchy management and adaptability to non-hierarchical dimensions. One of the possible inconveniences of a closed hierarchical datacube is that data consolidation may imply a loss of detail or an over-aggregation of data. By aggregating data at higher hierarchical levels, specific details may be lost at lower hierarchical levels, which can limit the ability to perform fine-grained data analysis at the lowest levels of granularity. That can be an inconvenience if details' analysis is needed for specific decisions or perspectives. A closed datacube can also be less flexible for \emph{ad hoc} analysis, because data are consolidated at specific hierarchical levels. If new analysis questions need different granularity levels of different data groupings, this may require to compute new closed datacubes or to recompute the current ones, which will require more time and resources. Also, closed datacubes are typically pre-computed to aggregate data at higher hierarchical levels, which can speed up analysis queries. However, this also means that data are pre-computed and stored in advance, which can incur a cost in terms of storage space for updates and modifications of underlying data. For example, hierarchies management can be complex in a closed datacube, because there can be more consolidation and aggregation levels to consider. The definition and management of hierarchical relations between dimensions can require a special attention to ensure that consolidations and aggregations are correctly aligned with the needs of the analysis. Finally, closed datacubes are made to work with hierarchical dimensions where there is a parent-child relation between granularity levels. However, they might not be as suitable for non-hierarchical dimensions, where data don't follow a clear hierarchical structure. In these cases, closed datacubes may not be as efficient to aggregate and consolidate data, which can lead to a loss in performance and accuracy in analysis results.


\section{Conclusion}

As part of multidimensional data warehouses (or datamart), after presenting an application case on a video game, having given a definition to the hierarchical dimensions and presented the different types of hierarchies (\cite{inmonBuildingDataWarehouse1996}), we have defined a theoretical framework and a formal definition of the attributes of hierarchical dimensions and their associated structure and hierarchy.

Inspired by the classical theory, and transposing it in a relatively direct way, we have defined, in a hierarchical context, the multidimensional space, specialisation orders (\cite{mitchellGeneralizationSearch1982}, \cite{mitchellMachineLearning1997}), functions and operators, in particular the Sum operator and the Product operator, and for each, defined at the dimensional level, overloaded and generalized on the cube lattice. 

We then characterized the hierarchical datacube lattice and gave a calculation of its size, showing the complexity of algorithms using the hierarchical cube lattice as a search space.

We then proposed the hierarchical datacube, which shows the same appeal as the classic datacube (\cite{grayDataCubeRelational1997}), major concept for data warehouse management.

As with the original datacube, the closed hierarchical datacube is its most concise representation. It exploits the generally large redundancies in the data. This is a good way to reduce the size of a datacube. By removing them, it allows the necessary size to be reduced substantially without losing useful data. When the data are very correlated, the representation by a closed hierarchical cube shows all its interest because the reduction brought is quite significant, and nevertheless makes it possible to find even more information than with a traditional datacube. These strong correlations appear in real data sets and increase when the dimensional data is very scattered over very large domains. Such characteristics are typical of data warehouses (\cite{rossFastComputationSparse1997}, \cite{beyerBottomUpComputationSparse1999}), and in particular in the field of video games where hierarchical dimension attributes are often legion.